\documentclass[preprint]{emulateapj}

\usepackage{color}
\usepackage{multirow}
\usepackage{rotating}
\usepackage{cancel,soul,ulem,amsmath} 

\def\hi{H{\sc i}}

\def\cm2{cm$^{-2}$}

\def\kms{km s$^{-1}$}

\def\nh3{NH$_3$}
\def\n2h{N$_2$H$^+$}

\def\13co{$^{13}$CO}
\def\c18o{C$^{18}$O}
\def\hc3n{HC$_3$N}
\def\h2{H$_2$}
\def\nh{n(H$_2$)}
\def\cp{C$^+$}
\def\lp{\>\> .}
\def\lc{\>\> ,}

\def\aj{AJ}
\def\aaps{A\&AS}
\def\aap{A\&A}
\def\apjs{ApJS}
\def\nat{Nature}
\def\apj{ApJ}

\newcommand{\Rmnum}[1]{\expandafter\@slowromancap\romannumeral #1@}
\slugcomment{ }
\usepackage[breaklinks,colorlinks,urlcolor=blue,citecolor=blue,linkcolor=blue]{hyperref}   
\makeatother

\newcommand{\noprint}[1]{}

\shorttitle{Where is OH and Does It Trace the Dark Molecular Gas (DMG)? }
\shortauthors{Li et al.}

\begin{document}

\title{Where is OH and Does It Trace the Dark Molecular Gas (DMG)? }

\author{Di Li\altaffilmark{1,2,3}, 
Ningyu Tang\altaffilmark{1},
Hiep Nguyen\altaffilmark{4,5},
J. R. Dawson\altaffilmark{4,5},
Carl Heiles\altaffilmark{6},
Duo Xu\altaffilmark{1,3,7},
Zhichen Pan\altaffilmark{1},
Paul F. Goldsmith\altaffilmark{8},
Steven J. Gibson \altaffilmark{9},
Claire E. Murray \altaffilmark{10,11},
Tim Robishaw \altaffilmark{12},
N.\ M.\ McClure-Griffiths \altaffilmark{13},
John Dickey \altaffilmark{14},
Jorge Pineda \altaffilmark{8},
Sne\v{z}ana Stanimirovi\'c \altaffilmark{10},
L. Bronfman\altaffilmark{15}, 
Thomas Troland\altaffilmark{16}, 
and the PRIMO collaboration\altaffilmark{17}
}

\altaffiltext{1}{National Astronomical Observatories, CAS, Beijing 100012, China; Email: dili@nao.cas.cn, nytang@nao.cas.cn}
\altaffiltext{2}{Key Laboratory of Radio Astronomy, Nanjing, Chinese Academy of Science}
\altaffiltext{3}{University of Chinese Academy of Sciences, Beijing 100049, China}
\altaffiltext{4}{Department of Physics and Astronomy and MQ Research Centre in Astronomy, Astrophysics and Astrophotonics, Macquarie 
 University, NSW 2109, Australia}
\altaffiltext{5}{Australia Telescope National Facility, CSIRO Astronomy and Space Science, PO Box 76, Epping, NSW 1710, Australia}
\altaffiltext{6}{Department of Astronomy, University of California, Berkeley, 601 Campbell Hall 3411, Berkeley, CA 94720-3411}
\altaffiltext{7}{Department of Astronomy, The University of Texas at Austin, Austin, TX 78712, USA}
\altaffiltext{8}{Jet Propulsion Laboratory, California Institute of Technology, 4800 Oak Grove Drive, Pasadena, CA 91109, USA}
\altaffiltext{9}{Western Kentucky University, Dept. of Physics and Astronomy, 1906 College Heights Blvd, Bowling Green, KY 42101, USA}
\altaffiltext{10}{University of Wisconsin, Department of Astronomy, 475 N Charter St., Madison, WI 53706, USA}
\altaffiltext{11}{Space Telescope Science Institute, 3700 San Martin Drive, Baltimore, MD 21218, USA}
\altaffiltext{12}{Dominion Radio Astrophysical Observatory, National Research Council, PO Box 248, Penticton,BC, V2A 6J9, Canada}
\altaffiltext{13}{Research School for Astronomy \& Astrophysics, Australian National University, Canberra, ACT 2611, Australia}
\altaffiltext{14}{University of Tasmania, School of Maths and Physics, Hobart, TAS 7001, Australia}
\altaffiltext{15}{Departamento de Astronom{\'{\i}}a, Universidad de Chile, Casilla 36, Santiago de Chile, Chile}
\altaffiltext{16}{Department of Physics and Astronomy, University of Kentucky, Lexington, Kentucky 40506}
\altaffiltext{17}{Pacific Rim Interstellar Matter Observers; http://ism.bao.ac.cn/primo}

\begin{abstract}
Hydroxyl (OH) is expected to be abundant in diffuse interstellar molecular gas as it forms along with \h2\ under similar conditions and within a similar extinction range. 
We have analyzed absorption measurements of  OH at 1665 MHz and 1667 MHz toward 44 extragalactic continuum sources,  together with the J=1-0  transitions  of $^{12}$CO, \13co, and C$^{18}$O, and the  J=2-1 of $^{12}$CO.   The excitation  temperature of OH were found to  follow a modified log-normal distribution
\begin{equation*}
f(T{\rm_{ex}}) \propto \frac{1}{ \sqrt{2\pi}\sigma } \rm{exp}\left[-\frac{[ln(\textit{T}_{ex})-ln(3.4\ K)]^2}{2\sigma^2}\right] \lc
\end{equation*}
 the peak of which is close to the temperature of the Galactic emission background (CMB+synchron). In fact, 90\% of the OH  has excitation temperature within 2 K of the Galactic background at the same location, providing a plausible explanation for the apparent difficulty to map  this abundant molecule in emission. The opacities of OH were found to be small and peak around 0.01.
For gas at intermediate extinctions  (A$\rm_V$\ $\sim$ 0.05--2 mag), the detection rate of OH with detection limit $N(\mathrm{OH})\simeq 10^{12}$ cm$^{-2}$  is approximately independent of $A\rm_V$. We conclude that OH is abundant in the diffuse molecular gas and OH absorption is a good tracer of `dark molecular gas (DMG)'.  The measured fraction of DMG depends on assumed detection threshold of the CO data set.  
The next generation of highly sensitive low frequency radio telescopes,  FAST and SKA, will make feasible the systematic inventory of diffuse molecular gas, through decomposing in velocity  the molecular  (e.g.\ OH and CH) absorption profiles toward background continuum sources with numbers exceeding what is currently available by orders of magnitude.

\end{abstract}


\section{Introduction}
 The two relatively denser phases of the interstellar medium (ISM) are the atomic Cold Neutral Medium (CNM) traced by the \hi\ $\lambda$ 21-cm hyperfine structure line and the `standard' molecular (H$_2$) clouds, usually traced by CO. CO has historically been the most important tracer of molecular hydrogen, which remains largely invisible due to its lack of emission at  temperatures in the molecular ISM. Empirically, CO intensities have been used as an indicator of the total molecular mass in the Milky Way and external galaxies through the so-called ``X-factor", with numerous  caveats, not least of which is the large opacities of CO transitions.  Gases in these two phases dominate the masses of star forming clouds on a galactic scale. The measured ISM gas mass from \hi\ and CO is thus the foundation of many key quantities in understanding galaxy evolution and star formation, such as the star formation efficiency. 
 
 A growing body of evidence, however,  indicates the existence of gas traced by neither \hi\ nor CO.  Comparative studies \citep[e.g.,][]{deVries1987} of Infrared Astronomy Satellite (IRAS) dust images and gas maps in \hi\ and CO revealed an apparent `excess' of dust emission. The \citet{Planck2011} clearly showed excess dust opacity  in the intermediate extinction range $A_V \sim$ 0.05--2 mag, roughly corresponding to the self-shielding thresholds of \h2\ and \13co, respectively. The missing gas, or rather, the undetected gas component, is widely referred to as dark gas, popularized as a common term by \citet{Grenier2005}. These authors found more diffuse gamma-ray emission observed by Energetic Gamma Ray Experiment Telescope (EGRET) than can be explained by cosmic-ray interactions with the observed H-nuclei. Observations of the THz fine structure \cp\ line also helped reveal the dark gas,  from which the \cp\ line strength  is stronger than can be produced by  only \hi\ gas \citep{Langer2010,Pineda2013,Langer2014}. A minority of the ISM community have argued that dark gas can be explained by underestimated H{\sc i} opacities \citep{Fukui2015}, which is in contrast with some other recent works \citep{Stanimirovic2014,Lee2015}.  We focus here on the dark molecular gas (DMG), or more specifically CO-dark molecular gas.

ISM chemistry and PDR models predict the existence of \h2\ in regions where CO is not detectable \citep{Wolfire2010}.
 CO can be of  low abundance due to photo-dissociation  in unshielded regions and/or can be heavily sub-thermal due to low collisional excitation rate in the diffuse gas.  OH, or Hydroxyl, was the first interstellar molecule detected at radio wavelenghs \citep{Weinreb1963}. It can form efficiently through relatively rapid routes including charge-exchange reactions  initiated by cosmic ray ionization once \h2\ becomes available \citep{vanDishoeck1988}. Starting from H$^+$, 
\begin{align}
& \mathrm{O+H^+\to O^+ + H} \lc \nonumber \\
& \mathrm{O^++H_2\to OH^+ + H}\ . \label{eq:ohplus} 
\end{align}
Also starting from H$_2^+$,
\begin{align}
& \mathrm{H_2^+ + H_2\to H_3^+ + H} \lc \nonumber \\
& \mathrm{H_3^+ + O \to OH^+ + H_2}\ . \label{eq:ohplus2} 
\end{align}
OH$^+$ then reacts with H$_2$ to form H$_2$O$^+$ that continues on to H$_3$O$^+$,
\begin{align}
& \mathrm{OH^+ + H_2 \to H_2O^+ + H} \lc \nonumber \\
& \mathrm{H_2O^+ + H_2 \to H_3O^+ + H}\ .\label{eq:h3op} 
\end{align}
H$_2$O$^+$ and H$_3$O$^+$ recombine with electrons to form OH.

OH can join the carbon reaction chain through reaction with C$^+$ and eventually  produce CO,
 \begin{align}
& \mathrm{OH+C^+\to CO^++ H} \lc \nonumber \\
& \mathrm{CO^++H_2 \to HCO^++ H} \lc  \nonumber \\
& \mathrm{HCO^++e^-\to CO+ H}\ .\label{eq:co}  
\end{align}
Widespread and abundant OH, along with HCO$^+$ and \cp\, is thus expected in diffuse and intermediate extinction regions. 

OH has  been widely detected throughout the Galactic plane \citep[e.g.,][]{Caswell1975,Turner1979,Boyce1994,Dawson2014,Bihr2015}, in local molecular clouds \citep[e.g.,][]{Sancisi1974,Wouterloot1985,Harju2000}, and in high-latitude translucent and cirrus clouds \citep[e.g.,][]{Grossmann1990,Barriault2010,Cotten2012}. Crucially, a small number of studies have confirmed OH extending outside CO-bright regions \citep{Wannier1993,Liszt1996,Allen2012,Allen2015}, and/or associated with narrow H{\sc i} absorption features \citep{Dickey1981,Liszt1996,Li2003}, confirming its viability as a dark gas tracer. OH and HCO$^{+}$ have been shown to be tightly correlated in absorption measurements against extragalactic continuum sources \citep{Liszt1996,Lucas1996}.

Because OH in emission is typically very weak, large-scale OH maps remain rare, particularly for diffuse molecular gas, which should presumably be dominated by DMG. Detectability often hinges on the presence of bright continuum background against which the OH lines can be seen in absorption, either the bright diffuse emission of the inner Galactic plane \citep[e.g.,][]{Dawson2014} or bright, compact extragalactic sources \citep[e.g.,][]{Goss1968,Nguyen-Q-Rieu1976,Crutcher1977,Crutcher1979,Dickey1981,Colgan1989,Liszt1996}. This latter approach has the additional advantage that on-source and off-source comparison can be made directly to derive optical depths and excitation temperatures. This is the approach we take in this work.

\citet{Heiles2003a,Heiles2003b} published the  Millennium Survey of 21-cm line absorption toward 79 continuum sources. The ON-OFF technique and Gaussian decomposition analysis allowed them to provide direct measurements of the excitation temperature and column density of \hi\ components. Given that the Millennium sources are generally out of the Plane, the absorption components are biased toward local gas. The large gain of Arecibo and the substantial integration time spent on each source made the Millennium Survey one of the most sensitive surveys of the diffuse ISM.
Among the significant findings is the fact that a substantial fraction of the CNM lies below the canonical 100K temperature predicted by phased ISM models \citep{Field1969,McKee1977} for maintaining pressure balance. 
The existence of cold gas in significant quantities points to the necessity of utilizing absorption measurements for a comprehensive census of ISM, taking into account the general Galactic radiation field. 

The L-wide receiver at Arecibo allows for simultaneous observation of \hi\ and OH. This was carried out by the Millennium Survey, but the OH data have remained unpublished until now. To analyze these OH absorption measurements in the context of DMG, we conducted 3mm and 1mm CO observations toward the Millennium sources and performed a combined analysis of their excitation and abundances.

This paper is organized as follows: In section \ref{sec:observations}, we describe the observations of \hi, OH, and CO. In section \ref{sec:oh_properties}, we analyze the OH line excitations and other properties. In section \ref{sec:comparison}, we explore the relation between these three spectral tracers. Discussions and conclusions are presented in section \ref{sec:abs_survey} and section \ref{sec:conclusion}, respectively. 

\section{Observations}
\label{sec:observations}

\subsection{\hi\ and OH}
\label{subsec:oh}
During the Millennium Survey, the $\Lambda$-doubling transitions of ground-state OH at 1665.402 and 1667.359 MHz were obtained simultaneously with \hi\ with the Arecibo L-wide receiver towards  
 72 of the 79 survey positions. These sources typically had flux density S$\rm_{1.4 GHz}$ $\gtrsim 2$ Jy, so produced antenna temperatures in excess of about 20 K at Arecibo.
The observations followed the strategy described by \citet{Heiles2001a} and \citet{Heiles2003a,Heiles2003b} -- the so-called Z17 method for obtaining and analyzing absorption spectra toward continuum sources. In this method, half of the integration time was spent on source, with the remaining time divided evenly among 16 OFF positions. The OFF spectra were then used to reconstruct the ``expected'' background gas spectrum, which would have been seen were there no continuum source present. The bandwidth of the OH observations was 0.78 MHz, with a channel width of 381 KHz, corresponding to a velocity resolution of 0.068 \kms. An RMS of 28 mK (T$_A$) per channel was achieved with 2 hours of total integration time. 
Twenty-one sightlines exhibited OH absorption. 

\begin{figure}
\includegraphics[width=1.0\linewidth]{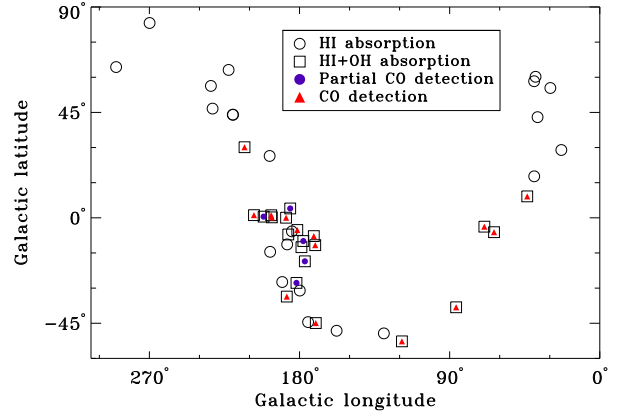}
\caption{The location of background continuum sources in Galactic coordinates.  This plot shows only the 44 sources towards which \hi\, OH and CO were observed. Open circles represent sources with detected \hi\ absorption only. Squares represent sources with detected \hi\ and OH absorption. Red triangles represent sources with CO detections, in which every detected OH component is also seen in CO. Blue dots represent sources with CO detections, in which some OH components do not  have corresponding CO detections. We call these sources ``partial CO detections".}\vspace{0.3cm}
\label{fig:posdistri} 
\end{figure}

\subsection{CO}
\label{subsec:CO}

We conducted a follow-up CO survey of 44 of the Millennium sight-lines for which OH data were taken.  Fig.~\ref{fig:posdistri} shows the distribution of all observed sources in Galactic coordinates.
    
	The J=1--0 transitions of CO, \13co, and \c18o\ were observed with the Purple Mountain Observatory Delingha (PMODLH) 13.7 m telescope of the Chinese Academy of Sciences. All numbers reported in this section are in the units of T$_\mathrm{mb}$, since the Delingha system automatically corrected for the main beam antenna efficiency. The three transitions were observed simultaneously with the 3 mm SIS receiver in March 2013, May 2013, May 2014, and May 2016. The FFTS wide-band spectral backend has a bandwidth of 1 GHz at a frequency resolution of 61.0 kHz, which corresponds to 0.159  \kms\ at 115.0 GHz and 0.166 \kms\ at 110.0 GHz. Position-switching mode was used with reference positions selected from the IRAS Sky Survey Atlas\footnote{$http://irsa.ipac.caltech.edu/data/ISSA/$}.  The system temperature varied from 210 K to 350 K for CO, and 140 K  to 225 K for the \13co and C$^{18}$O observations. The resulting RMS are $\sim$ 60 mK for a 0.159 \kms\ channel for $^{12}$CO and $\sim$ 30 mK for a 0.166 \kms\ channel for $^{13}$CO and C$^{18}$O, respectively.

	The $^{12}$CO(J=2--1) data were taken with the  Caltech Submillimeter Observatory (CSO) 10.4 m on Mauna Kea in July, October, and December of 2013.  The system temperature varied from 230 to 300 K for $^{12}$CO(J=2--1), resulting in an RMS of $\sim 35$ mK at a velocity resolution of 0.16 \kms. 
	
To achieve consistent sensitivity among sight-lines, $^{12}$CO(J=2--1) spectra toward 3 sources were also obtained with the IRAM 30m telescope in frequency-switching mode on the 22nd and 23rd of May, 2016.  The integration times of these observation were between 30 and 90 minutes, resulting in an RMS of less than 20 mK at a velocity resolution of 0.25 \kms.
	
	The astronomical software package Gildas/CLASS\footnote{$http://www.iram.fr/IRAMFR/GILDAS/$} was used for data reduction including baseline removal and Gaussian fitting.

\section{OH Properties}
\label{sec:oh_properties}

\subsection{Radiative Transfer and Gaussian Analysis}
\label{sec:oh_emi_abs}

\label{sec:TE}
The equations of radiative transfer for ON/OFF source measurements may be written as
\begin{align}
& T_\mathrm{A}^\mathrm{ON} (v)/\eta_{b} =(T_\mathrm{ex}-T_\mathrm{bg}-T_\mathrm{c})(1-e^{-\tau_{v}})\ \ \ \ \mathrm{(K)}\ \label{eq4a_simp}, \\
& T_\mathrm{A}^\mathrm{OFF} (v)/\eta_{b}=(T_\mathrm{ex}-T_\mathrm{bg})(1-e^{-\tau_{v}})\ \ \ \ \mathrm{(K)} \label{eq4b_simp} \lc
\end{align}
where we assumed main beam efficiency $\eta_{b} = 0.5$ according to \citet{Heiles2001b} 
in all subsequent calculations. $T\rm_{ex}$ and $\tau_v$ are excitation temperature and optical depth of the cold cloud, respectively. $T_\mathrm{A}^\mathrm{ON} (v)$ and $T_\mathrm{A}^\mathrm{OFF} (v)$ are antenna temperatures toward and offset from the continuum source, respectively.  Here, $T_\mathrm{c}$ is the compact continuum source brightness temperature, and $T_\mathrm{bg}$ is the background brightness temperature, consisting of the 2.7 K isotropic CMB and the Galactic synchrotron background at the source position. We adopted the same treatment of the background continuum \citep{Heiles2003a}  
\begin{equation}
T{\rm_{bg}} = 2.7 + T{\rm_{bg408}}(\nu{\rm_{OH}}/408\ \rm MHz)^{-2.8} \lc
\label{eq4c} 
\end{equation}
where $T_{bg408}$ comes from \citet{Haslam1982}.
The  background continuum contribution from Galactic H{\sc ii} regions can be safely ignored, since our sources are either at high Galactic latitudes or Galactic Anti-Center longitudes. Typical $T\rm_{bg}$ values are thus found to be around 3.3 K. 

We decomposed the OH spectra into Gaussian components to evaluate the physical properties of OH clouds along the line-of-sight. Following the methodology of \citet{Heiles2003a}, we assumed a two-phase medium, in which cold gas components are seen in both absorption and emission (i.e.\ in both the opacity and brightness temperature profiles), while warm gas appears only in emission, i.e.\ only in brightness temperature (see \citet{Heiles2003a} for further details).  While this technique is generally applicable for both H{\sc i} and OH, we have only detected OH in absorption in  this work. ``Warm'' OH components in emission have been observed by \citet{Liszt1996}, although only by the NRAO 43m and not with the VLA or Nancay in their study. This is consistent with our results in that there is no OH warm enough ($>$ a few hundred K) in our Arecibo beam to be seen in emission, nor do we expect it from astrochemistry considerations. 

In brief, the expected profile $T_\mathrm{exp}(v)$ consists of both emission and absorption components:
\begin{equation}
T_\mathrm{exp}(v) = T_\mathrm{B,cold}(v) + T_\mathrm{B,warm}(v) \lc
\label{eq4d} 
\end{equation}
\noindent where $T_\mathrm{B,cold}(v)$ is the brightness temperature of the cold gas and $T_\mathrm{B,warm}(v)$ is the brightness temperature of the the warm gas. Both components contribute to the emission profile. 

The opacity spectrum is obtained by combining the on- and off-source spectra (equations \ref{eq4a_simp} and \ref{eq4b_simp}), and contains only cold gas seen in absorption. First, we fit the observed opacity spectrum $e^{-\tau(v)}$ with a set of $N$ Gaussian components. 

Next we fit the expected emission profile, $T_\mathrm{exp}$, which is assumed to also consist of the $N$ cold components seen in the absorption spectrum, plus any warm components seen only in emission. We further assume that each component is independent and isothermal with an excitation temperature $T_{ex,n}$:
\begin{equation}
\tau(v) = \sum\limits_{n=0}^{N-1} \tau_{0,n}e^{-[(v-v_{0,n})/\delta v_{n}]^2}. 
\label{eqtau} 
\end{equation}
Here $\tau_{0,n}$, $v_{0,n}$, $\delta v_{n}$ are respectively the peak optical depth, central $V\rm_{LSR}$, and 1/$e$-width of component $n$. All the values of $\tau_{0,n}$, $v_{0,n}$ and $\delta v_{n}$ were then obtained through least-squares fitting. The contribution of the cold components is given by
\begin{equation}
T_\mathrm{B,cold}(v) = \sum\limits_{n=0}^{N-1} T_{\mathrm{ex},n}(1-e^{-\tau_{n}(v)}) e^{-\sum\limits_{m=0}^{n-1} \tau_{m}(v)}\lc
\label{eqtbc} 
\end{equation}
where subscript $m$ describes  the $M$ absorbing clouds lying in front of the $n^{th}$ cloud. When n=0, the summation over m takes no effect, as there is no foreground cloud.

We obtained the values of excitation temperature $T_{\mathrm{ex},n}$ from this fit. As in \citet{Heiles2003a}, we experimented with all possible orders along the line of sight and retained the one that yields the smallest residuals.

\begin{figure}
\includegraphics[width=1.0\linewidth]{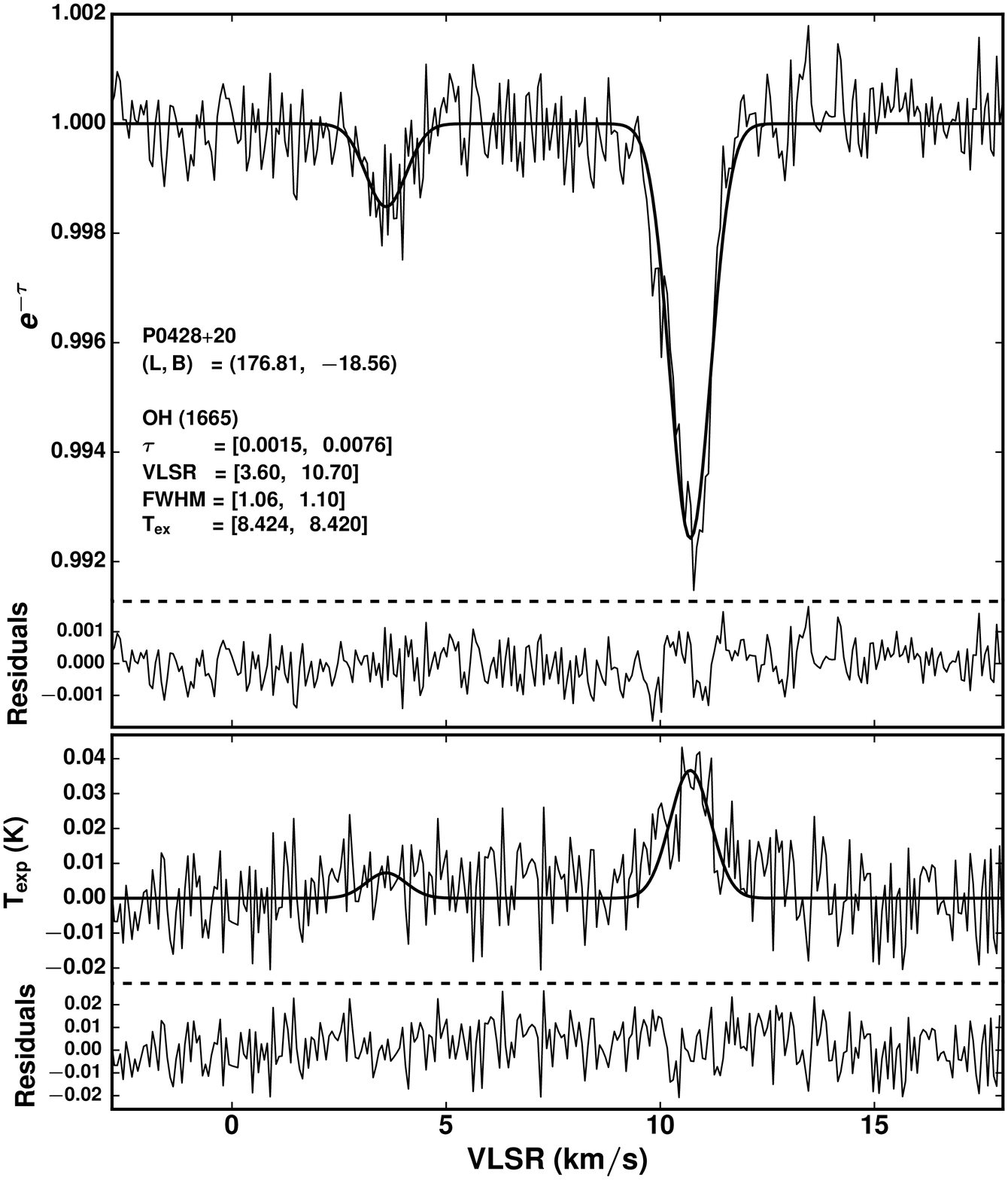}
\caption{Illustration of derived parameters from the fits to the absorption (upper panel) and expected emission (lower panel) profiles for the source P0428+20. The thin solid lines show the data; the thick solid lines are the fits. The solid lines below the dashed-lines present the residuals for absorption and fitted expected emission, respectively.}
\label{fig:p042820.eps}
\end{figure}

\begin{figure}
\includegraphics[width=1.0\linewidth]{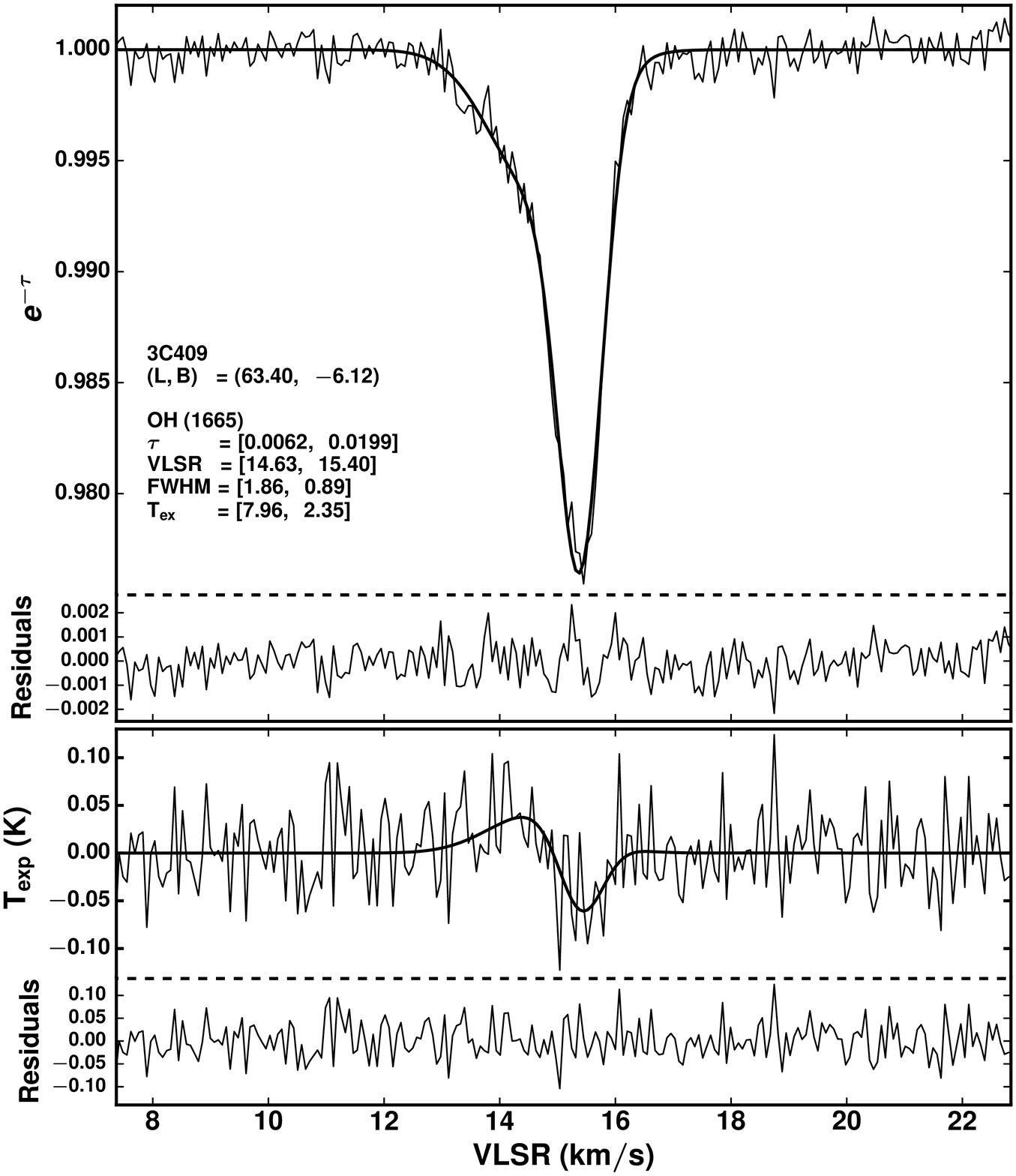}
\caption{Data for the source 3C409. See Fig.2 for complete description.}
\label{fig:3c409.eps}
\end{figure}

In total, we detected 48 OH components towards 21 sightlines. Example spectra and fits are shown in Figures \ref{fig:p042820.eps} and \ref{fig:3c409.eps}.%

\begin{figure}
\includegraphics[width=1.0\linewidth]{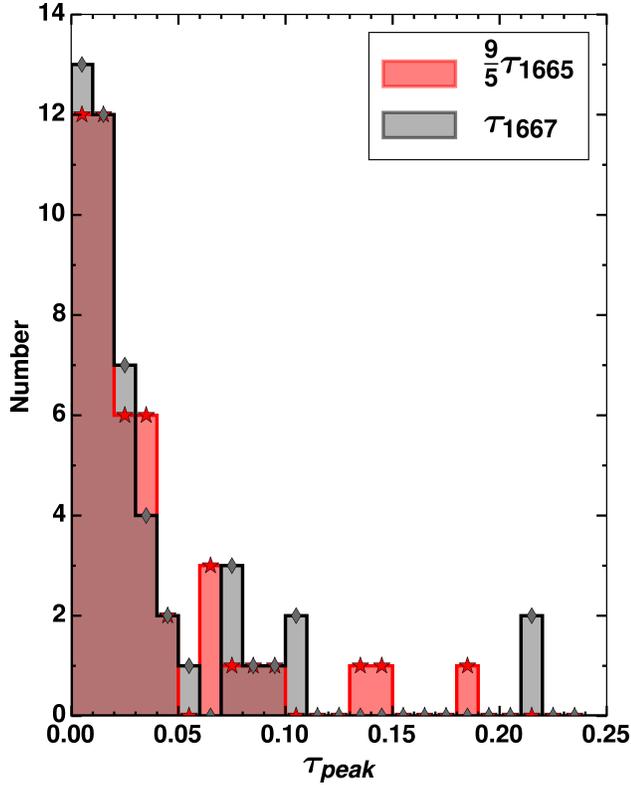}
\caption{Histogram of peak optical depths of the OH Gaussian components. Black shows the results for the 1667 MHz line and red shows 1.8 times the results for the 1665 MHz line.}
\label{fig:taupeak_hist} 
\end{figure}

\begin{figure}
\includegraphics[width=1.0\linewidth]{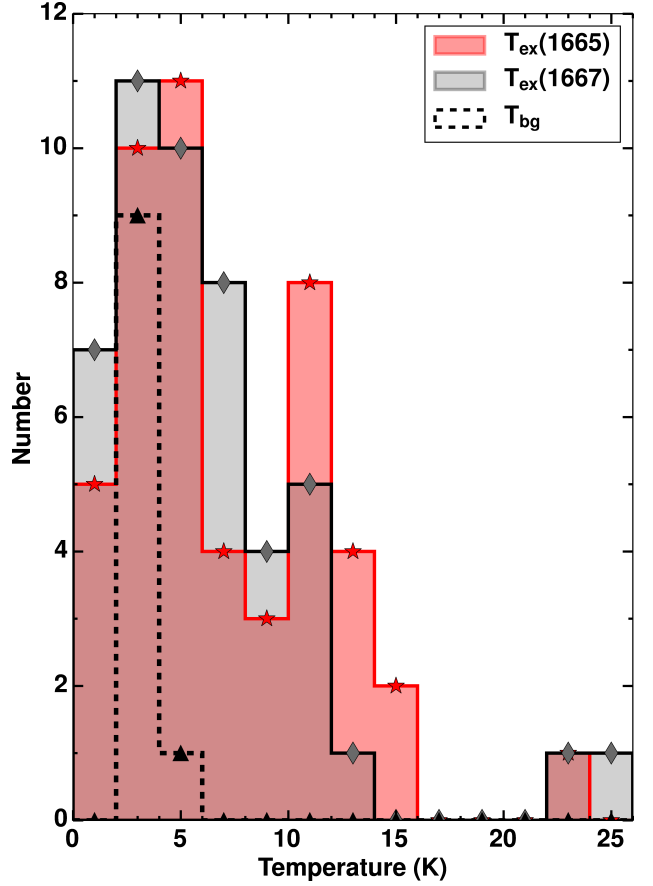}
\caption{Histograms of excitation temperature, $T\rm_{ex}$, of the Gaussian components for the two OH main lines and the background continuum temperature $T\rm_{bg}$.  The number histogram of $T\rm_{bg}$ has been scaled by a factor of 0.5, for ease of visualization.}
\label{fig:Tex_hist} 
\end{figure}

\subsection{OH Excitation and Optical Depth}
\label{subsec:oh_excitation}
Our fitting scheme  provides measurements of the excitation conditions and optical depths of the OH gas. 

Figure \ref{fig:taupeak_hist} shows a histogram of optical depths in the two OH main lines, with the 1665 line scaled by a factor of 9/5 (see below). The measured optical depth of both lines peaks at very low values of $\sim0.01$, with a longer tail extending as high as 0.21 in the 1667 line. As can be seen in Table \ref{table:1}, the uncertainties on these values are generally very small. The OH gas probed by absorption is thus quite optically thin. 

\begin{figure}
\includegraphics[width=1.0\linewidth]{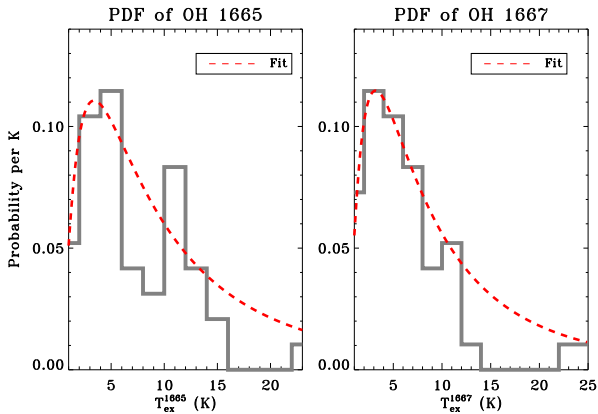}
\caption{Fitting results of the PDFs of OH excitation temperature for both the 1665 and 1667 lines. Fit parameters are given in Table \ref{table:Tex_hist_fit}.} 
\label{fig:Tex_hist_fit} 
\end{figure}

As seen in Figure \ref{fig:Tex_hist}, OH excitation temperatures in the two main lines peak at $\sim3$--4 K and are similar for the two lines. The majority ($\sim90\%$) of the components show $T_\mathrm{ex}$ values within 2 K of the diffuse continuum background temperature, $T_\mathrm{bg}$, consistent with measurements from past work \citep[e.g.,][]{Nguyen-Q-Rieu1976,Crutcher1979,Dickey1981}. The probability density distribution (PDF) of the OH excitation temperature can be fitted by  a modified normalized lognormal function\footnote{Equation \ref{eq:lognormal} differs from standard log-normal function in that it has no variable (x) in the denominator.}, 
\begin{equation}
f(T_\mathrm{ex}) \propto \frac{1}{ \sqrt{2\pi}\sigma} \mathrm{exp}\left[-\frac{[\mathrm{ln}(T_\mathrm{ex})-\mathrm{ln}(T^0_\mathrm{ex})]^2}{2\sigma^2}\right] \lp
\label{eq:lognormal}
\end{equation}
 The fit parameters are sensitive to the binning and statistical weighting of the data and are not unique. We chose one solution that preserves the tail of relatively high $T_\mathrm{ex}$. The exact numerical values are less meaningful than the location of the distribution peak and the rough trend of $T_\mathrm{ex}$, which are represented in the current fitting. 
 The fit results are shown in Figure \ref{fig:Tex_hist_fit} and Table \ref{table:Tex_hist_fit}.  The statistical uncertainties are small. Similar numerical values are found in fitting both the 1665 and 1667 transitions. A simple average of the two sets of fit parameters is also presented in Table \ref{table:Tex_hist_fit}. Given the unbiased nature of absorption-selected sightlines and the fact that almost half of the detected OH components lie at low Galactic Latitudes ($|b|< 5^\circ$), such a generic distribution function \sout{can}  may be representative of OH excitation conditions in the Galactic ISM. 
 This tentative conclusion will be refined by future observations. 

\begin{deluxetable}{lcccc}
\tablewidth{0pt}
\tablecaption{Lognormal fit parameters for the OH $T_\mathrm{ex}$ distribution, as shown in Figure \ref{fig:Tex_hist_fit}. \label{table:Tex_hist_fit}}
\tablehead{
 Line     & Fitted $T^0_\mathrm{ex}$$^a$  & Fitted $\sigma$$^a$
 }
\startdata
 OH 1665 &  3.4 & 0.98   \\   
 OH 1667 &  3.2 & 0.96	\\
 OH average &  3.3 & 0.97
 \enddata
 \tablenotetext{a} {Parameter defined in Equation \ref{eq:lognormal}.}
\end{deluxetable}

 The combined results of $T_\mathrm{ex}$ and $\tau$ reflect the complexities in OH excitation.
When the level populations of the OH ground states are in LTE, the excitation  temperatures of the 1665 and 1667 MHz lines are equal, and their optical depth ratio ($R_{67/65}=\tau_{1667}/\tau_{1665}$) is 1.8. In general, however, we do not expect OH to be thermally excited. The satellite lines (at 1612 and 1720 MHz) commonly show highly anomalous excitation patterns, with $T_\mathrm{ex}$ that are strongly subthermal in one line, and either very high or negative (masing) in the other \citep[e.g.,][]{Dawson2014}. This may occur due to far-IR or infrared pumping, which can lead to selective overpopulation of either the $F=1$ or the $F=2$ level pair \citep[e.g.,][]{Guibert1978}. The result is that the main line $T_\mathrm{ex}$ may be very similar, despite the system as a whole being strongly non-thermally excited.

\begin{figure}
\includegraphics[width=1.0\linewidth]{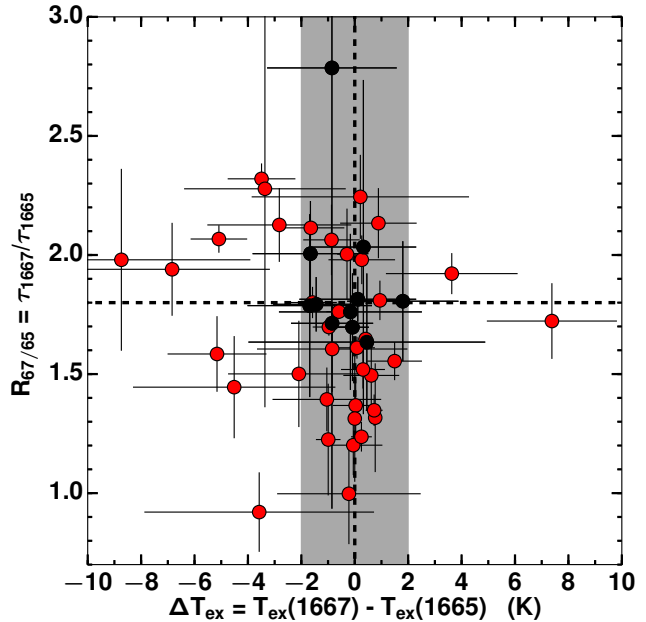}
\caption{Optical depth ratio ($R_{67/65}=\tau_{1667}/\tau_{1665}$) as a function of  excitation temperature difference ($T_\mathrm{ex}(1667)-T_\mathrm{ex}(1665)$) for the 1667 and 1665 MHz OH lines. The horizontal and vertical dashed lines indicate 1.8 and 0.0, the values for LTE excitation. Error bars indicate the $1\sigma$ formal uncertainties propagated through from the Gaussian fits. Black points are those consistent with LTE to within the $1\sigma$ errors; red points are inconsistent. The vertical shaded region represents $|\Delta T\rm_{ex}| < 2$ K.}
\label{fig:delspin_vs_tauratio} 
\end{figure}

As shown in Figure \ref{fig:delspin_vs_tauratio}, most OH components deviate from LTE at more than the formal $1\sigma$ uncertainties propagated  from the Gaussian fits. The difference between the excitation temperatures, however, mainly falls within 2 K ($|\Delta T\rm_{ex}| < 2$ K), consistent with values observed in previous absorption measurements toward continuum sources \citep[e.g.,][]{Nguyen-Q-Rieu1976,Crutcher1977,Crutcher1979,Dickey1981}. 

\subsection{OH Column Density}
\label{subsec:oh_column}
	The column densities of the OH components were computed independently from each line according to:
\begin{equation}
N(\mathrm{OH})_{1667}=\frac{8\pi k~T_{\mathrm{ex},1667}~{{\nu}^2_{1667}}}{A_{1667}~c^{3}h}\frac{16}{5}\int \tau_{1667} \,dv \lc
\label{5a}
\end{equation}
\begin{equation}
N(\mathrm{OH})_{1665}=\frac{8\pi k~T_{\mathrm{ex},1665}~{{\nu}^2_{1665}}}{A_{1665}~c^{3}h}\frac{16}{3}\int \tau_{1665} \,dv \lc
\label{eq5b}
\end{equation}
where $A_{1667}=7.778\times 10^{-11}$ $s^{-1}$ and $A_{1665}=7.177\times 10^{-11} s^{-1}$ are the Einstein A-coefficients of the OH main lines \citep{Destombes1977}. The values are tabulated in Table \ref{table:1} and discussed below. 

\section{The Relation Between H{\sc i}, CO, and OH}
\label{sec:comparison}
\subsection{OH and \hi\ }
\label{subsec:hioh}
\begin{figure}
\includegraphics[width=1.0\linewidth]{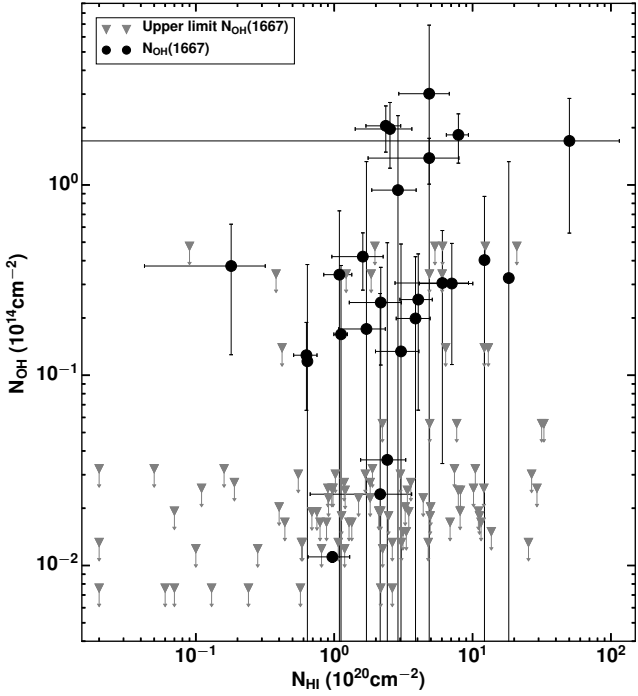}
\caption{OH column densities derived from the 1667 line, $N(\mathrm{OH})$, versus H{\sc i} column densities, $N(\mathrm{HI})$, both on a Gaussian component-by-component basis. The triangles depict approximate upper limits for \hi\ components where OH was not detected (see Section \ref{subsec:hioh}).}
\label{fig:Noh_vs_NHI} 
\end{figure}

Regardless of the complexities in the OH excitation, the measured line ratios  are generally close to the LTE value of 5/9 with a relatively small scatter. To appropriately utilize both transitions while minimizing the impact of the poorer S/N in the 1665 line, we estimate the total OH column density as $N(\mathrm{OH})=\frac{5}{14} N(1665)+ \frac{9}{14} N(1667)$.
The OH column densities obtained with this method are plotted against \hi\ column density of each Gaussian component in Figure \ref{fig:Noh_vs_NHI} on a  component-by-component basis. Since the \hi\ components are always wider than OH lines, the sum of all OH components was used when multiple OH components coincide with one \hi\ component. 

For non-detections, an upper limit was estimated based on the 1667 MHz spectrum alone, assuming a single Gaussian optical depth spectrum, with a FWHM of 1.0 km/s and peak $\tau$ equal to 3 times the spectral RMS.  For instance, rms of 1667 MHz absorption spectrum toward 3C138 is 31 mK in brightness temperature. $T_\mathrm{ex}$ was assumed to be equal to 3.5 K, the peak of the lognormal function fitted to the $T_\mathrm{ex}$ distribution in Figure \ref{fig:Tex_hist_fit}. These values are plotted as triangles in Figure \ref{fig:Noh_vs_NHI}. Our absorption data set typically has a detection limit around $N(\mathrm{OH})\simeq 10^{12}$ cm$^{-2}$, with a number of higher limits occurring towards weaker continuum background sources.

Many \hi\ components have no detectable OH. Where OH is detected, there is some suggestion of a weak correlation between the OH and \hi\ column densities. For $N(\mathrm{HI})$ between 10$^{20}$ and 10$^{21}$ cm$^{-2}$ most sources are consistent with an [OH]/[HI] abundance ratio $\sim$10$^{-7}$.

\subsection{OH and CO}
\label{subsec:ohco}

Both CO and OH are widely used molecular tracers in the ISM. \citet{Allen2012,Allen2015} performed a pilot OH survey toward the Galactic plane around $l\approx 105^{\circ}$ and demonstrated the presence of CO-dark molecular gas (DMG): CO emission was absent in more than half of the detected OH spectral features  (see also \citet{Wannier1993} and \citet{Barriault2010}). \citet{Xu2016} took OH observations across a boundary of the Taurus molecular cloud, revealing that the fraction of DMG decreases from 0.8 in the outer CO poor region to 0.2 in the inner CO abundant region.

Our explicit measurements of CO and OH properties allow us to examine the relation between CO and OH again. Nine OH  components are identified as DMG clouds and will be discussed in detail in section \ref{subsec:darkgas}. We focus here on molecular clouds with both OH and CO detections. 

CO emission was detected toward 40/49
OH components.  CO column density, $N$(CO), is calculated  differently for each of the following three cases:

\begin{enumerate}
\item   Detection of only $^{12}$CO(J=1--0).
\item   Detection of both $^{12}$CO(J=1--0) and $^{12}$CO(J=2--1).
\item   Detection of three lines, $^{12}$CO(J=1--0), $^{12}$CO(J=2--1), and $^{13}$CO(J=1--0) simultaneously.
\end{enumerate}

In LTE, the total column density $N\rm_{tot}$ of a two-level transition from upper level $u$ to lower level $l$ is given by
\begin{equation}
N_{\textrm{tot}}=\frac{8\pi\nu^3}{c^3A_{ul}}\frac{Q_{rot}}{g_u}\frac{e^{E_u/kT_\textrm{ex}}}{e^{h\nu/kT{\rm_{ex}}}-1} \int \tau_{\upsilon} d\upsilon
\label{eq:col}
\end{equation} 
where $A_{ul}$ is the spontaneous emission coefficient for transitions between levels $u$ and $l$, $g_{u}$ is the degeneracy of level $u$, $E_u$ is the energy of level $u$, and $T\rm_{ex}$ is the excitation temperature. $Q\rm_{rot}$ is the rotational partition function, given by 

\begin{align}
Q_\textrm{rot}  \approx  \frac{kT_\textrm{ex}}{hB_0}+\frac{1}{3}+\frac{1}{15}\left(\frac{hB_0}{kT_\textrm{ex}}\right) & +\frac{4}{315}\left(\frac{hB_0}{kT_\textrm{ex}}\right)^2 \nonumber \\ 
& +\frac{1}{315}\left(\frac{hB_0}{kT_\textrm{ex}}\right)^3 ,
\label{eq:roteq}
\end{align}
which is good to $<1$\% when $T\rm_{ex}> 2$ K. $B_0$ is the rotational constant of the molecule.

$\tau_{\nu}$ is the optical depth of the line, and is related to brightness temperature, $T_\mathrm{b}$, 
through the radiative transfer equation

\begin{equation}
J(T_\mathrm{b}) = f[J(T_\mathrm{ex})-J(T_\mathrm{bg})][1-e^{-\tau_{v}}]
 \label{eq:rtran}
 \end{equation}
where $f$ is the beam filling factor and $J(T)=(h\nu/k)/(e^{h\nu/kT}-1)$.

When only $^{12}$CO(J=1--0) is detected (case 1), we do not adopt the optically thick assumption but adopt the assumption that the excitation temperature of $^{12}$CO is 10 K as suggested by \citet{Goldsmith2008} in the Taurus cloud.  Optical depth $\tau_v$ can be derived from equation \ref{eq:rtran}. Then the total column density of $^{12}$CO can be derived through combination of equations \ref{eq:col} and \ref{eq:roteq}. 

For case 2, we adopted the assumption of optically thin $^{12}$CO(J=1--0) and $^{12}$CO(J=2--1) lines. This is reasonable since there is no $^{13}$CO(J=1--0) detection and the corrected antenna temperature of CO are smaller than 1 K (see e.g.\ source 3C105 in Fig.\ \ref{fig:spec1}). The excitation temperature of $^{12}$CO, $T\rm_{ex}$, is then obtained from the following equation 

\begin{multline}
\frac{e^{11.06/T_\textrm{bg}-16.59/T_\textrm{ex}}-e^{-5.53/T_\textrm{ex}}}{e^{5.53/T_\textrm{bg}-5.53/T_\textrm{ex}}-1} = \frac{\nu_{21}^3A_{10}g_1}{2\nu_{10}^3A_{21}g_2}\times \\
\left[\frac{e^{11.06/T_\mathrm{bg}}-1}{e^{5.53/T_\textrm{bg}}-1}\right]\frac{\int T^{21}_\mathrm{R} d\upsilon}{\int T^{10}_\mathrm{R} d\upsilon},
\label{eq:tspinco}
\end{multline}
where $T_\mathrm{R} = J(T_\mathrm{b})$. Once $T\rm_{ex}$ in equation \ref{eq:tspinco} is determined, $N(^{12}$CO) is derived as that in case 1. 

When both $^{12}$CO(J=1--0) and $^{13}$CO(J=1--0) are detected as in case 3, we assume $\tau_{10}(\rm ^{12}CO)\gg 1$ and derive $T\rm_{ex}$ and $\tau_{10}(\rm ^{13}CO)$ via:

\begin{equation}
 T_\textrm{ex}=5.53\left\{\rm ln\left[1+\frac{5.532}{\textit{T}_R(^{12}CO)+0.819}\right]  \right\}^{-1}.  
\label{eq:tspin13}
\end{equation}

\begin{multline}
\tau_{10}(\rm ^{13}CO) = -ln\{1-\frac{\textit{T}_R(^{13}CO)}{5.29}(\left[\textit{e}^{5.29/\textit{T}_{ex}}-1\right]^{-1} \\
  -0.16)^{-1}\}.
\label{eq:tau13}
\end{multline}

$N\rm_{tot}(^{13}CO)$ is then obtained using equation \ref{eq:roteq},  from which we compute $N\rm_{tot}(^{12}CO)$ by assuming $\rm [^{12}C]/[^{13}C]$=68 \citep{Milam2005}.

\begin{figure}[htp]
\includegraphics[width=0.98\linewidth]{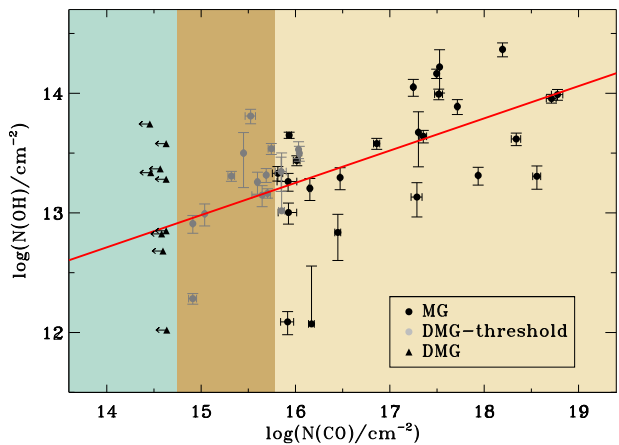}
\caption{CO column densities, $N$(CO), versus OH column densities, $N(\mathrm{OH})$ for three categories of clouds. Molecular clouds (MG) with CO detections are represented with black filled circles in the light yellow region. DMG-threshold clouds  are represented by grey filled circles in the dark yellow region. In these clouds, CO emission is detected at the CO sensitivity of this work but would be detected at less than 3$\sigma$ under a CO sensitivity of 0.25 K per 0.65 km s$^{-1}$, typical of the CFA CO survey \citep{Dame2001}. Black triangles in the blue region represent DMG clouds in which CO emission is not detected at the CO sensitivity of the present work. The red solid line represents the least-squares fit result to the MG and DMG-threshold clouds.}
\label{fig:co_vs_oh} 
\end{figure}

 The central velocities of derived OH components are used as initial estimates for CO fitting with the CLASS/GILDAS software. The properties of the fitted CO components are shown in Table \ref{table:2}, and the above calculations of CO excitation temperature and total column density are based on the derived Gaussian fitting results.
The coincidence between OH and CO components is judged by central velocity. The velocity difference should be less than 0.5 km s$^{-1}$. We plot the relation between OH and CO column density in Figure \ref{fig:co_vs_oh}. A positive correlation between log(N(OH)) and log(N(CO)) is seen.  Least-squares fitting yields the relation log(N(OH)) = 8.16$^{0.898}_{0.898}$+ 0.316$^{0.054}_{0.054}$ log(N(CO)). The linear Pearson correlation coefficient is 0.69, indicating a strong correlation. This result is consistent with \citet{Allen2015}, where an apparent correlation was found between strong $^{12}$CO emission and  OH emission.

\subsection{CO-Dark Molecular Gas}
\label{subsec:darkgas}

All-sky CFA CO survey data \citep{Dame2001} have been widely used to investigate the distribution of CO on large scales. \citet{Planck2011}, for example, analyzed Planck data along with the CFA CO survey to probe the large-scale DMG distribution. In principle, the definition of a DMG cloud depends on the sensitivity of the CO data employed. For example, \citet{Donate2017} found that the fraction of CO-dark molecular gas relative to total H$_2$ decreased from 58\% to 30\% in the Pegasus-Pisces region when higher sensitivity CO data were taken. In this study, the representative 1-$\sigma$ sensitivity of CFA CO data is about $\sigma\rm_{CFA}$=0.25 K per 0.65 km s$^{-1}$. Using this as an illustrative detection threshold, we here identify components as ``DMG-threshold clouds'' in cases where CO emission would be undetected at 3$\sigma\rm_{CFA}$ but is detected at the higher sensitivity of the present work.

\begin{figure*}[htp]
\includegraphics[width=0.34\linewidth]{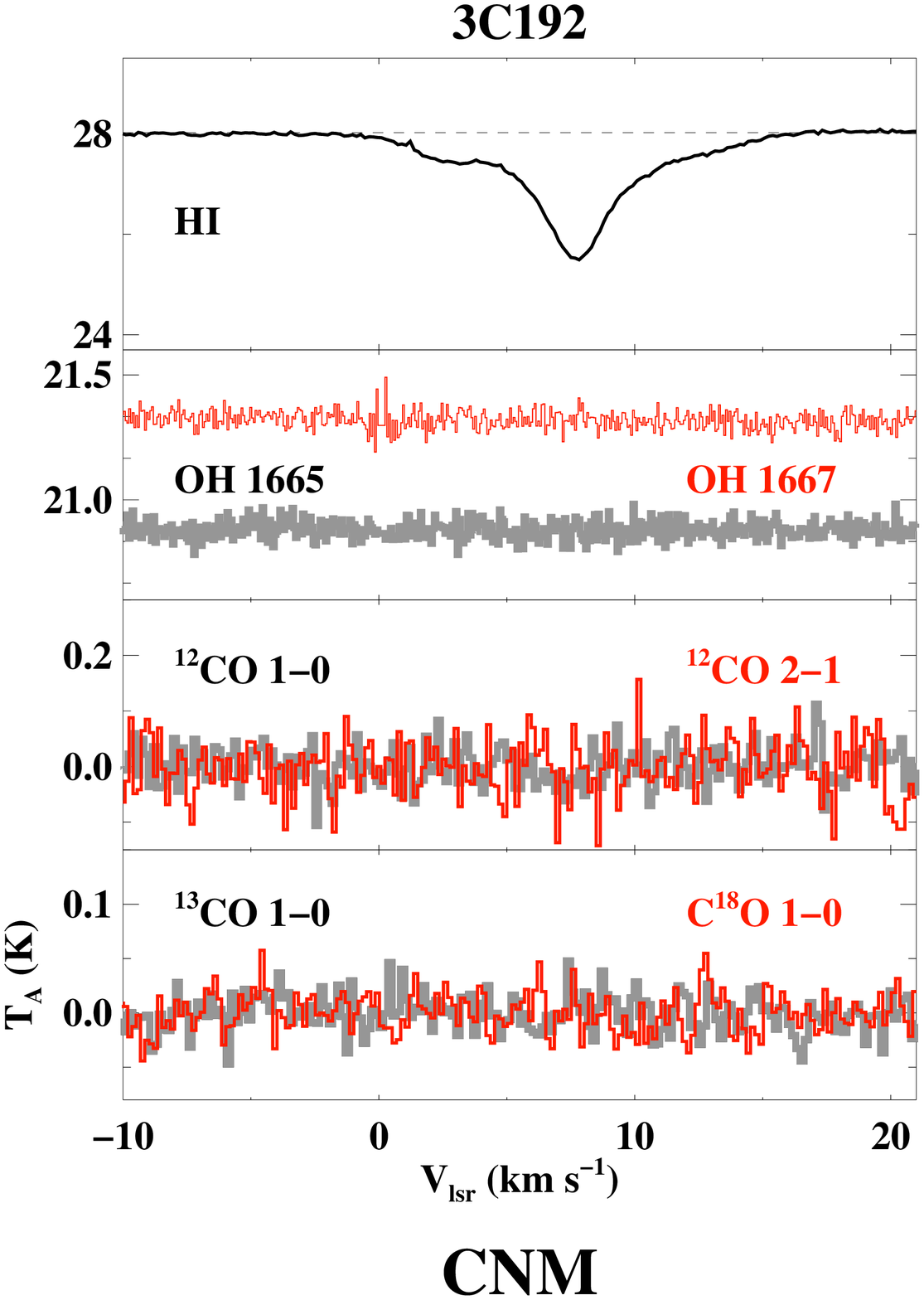}
\includegraphics[width=0.34\linewidth]{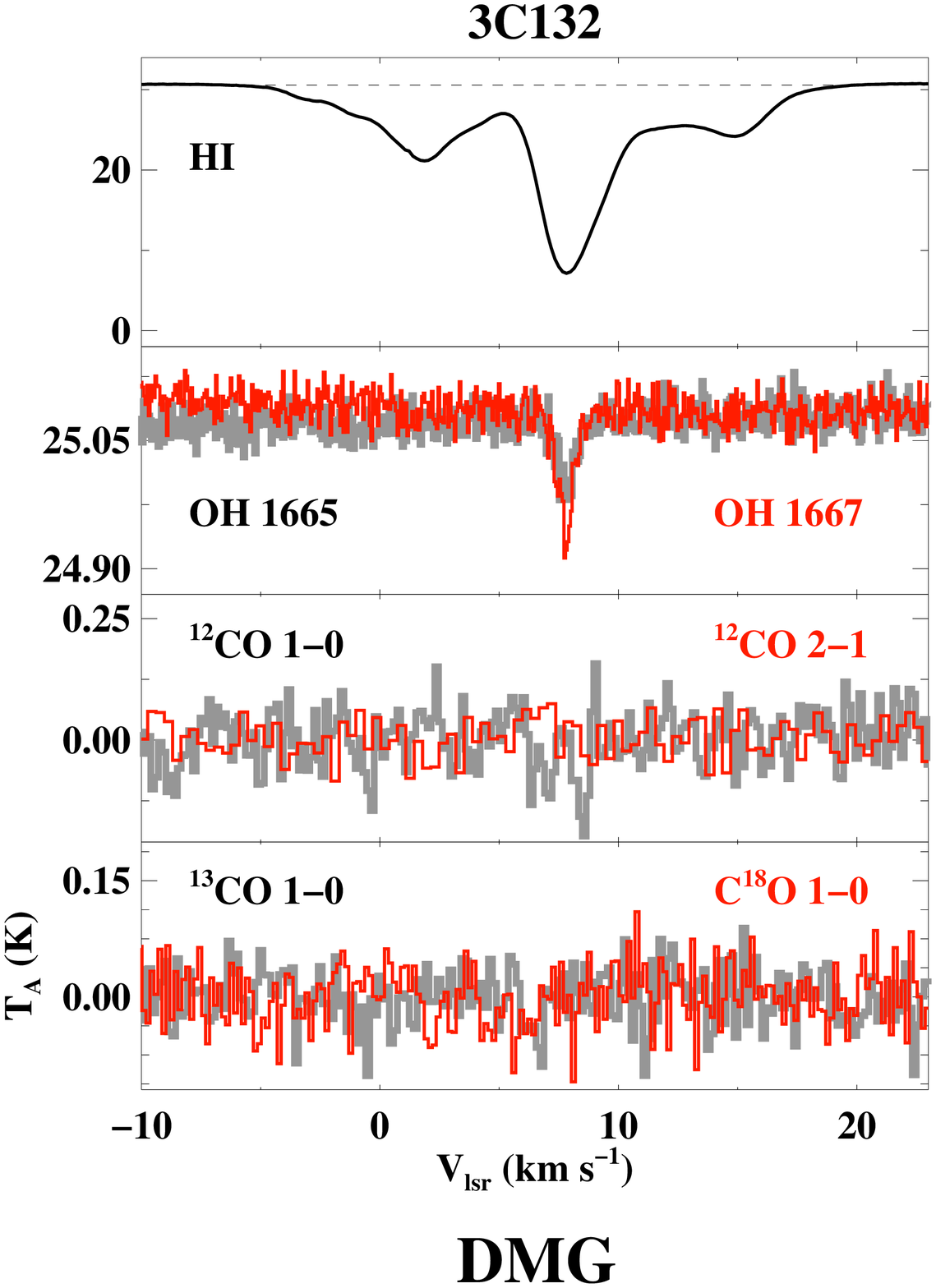}
\includegraphics[width=0.34\linewidth]{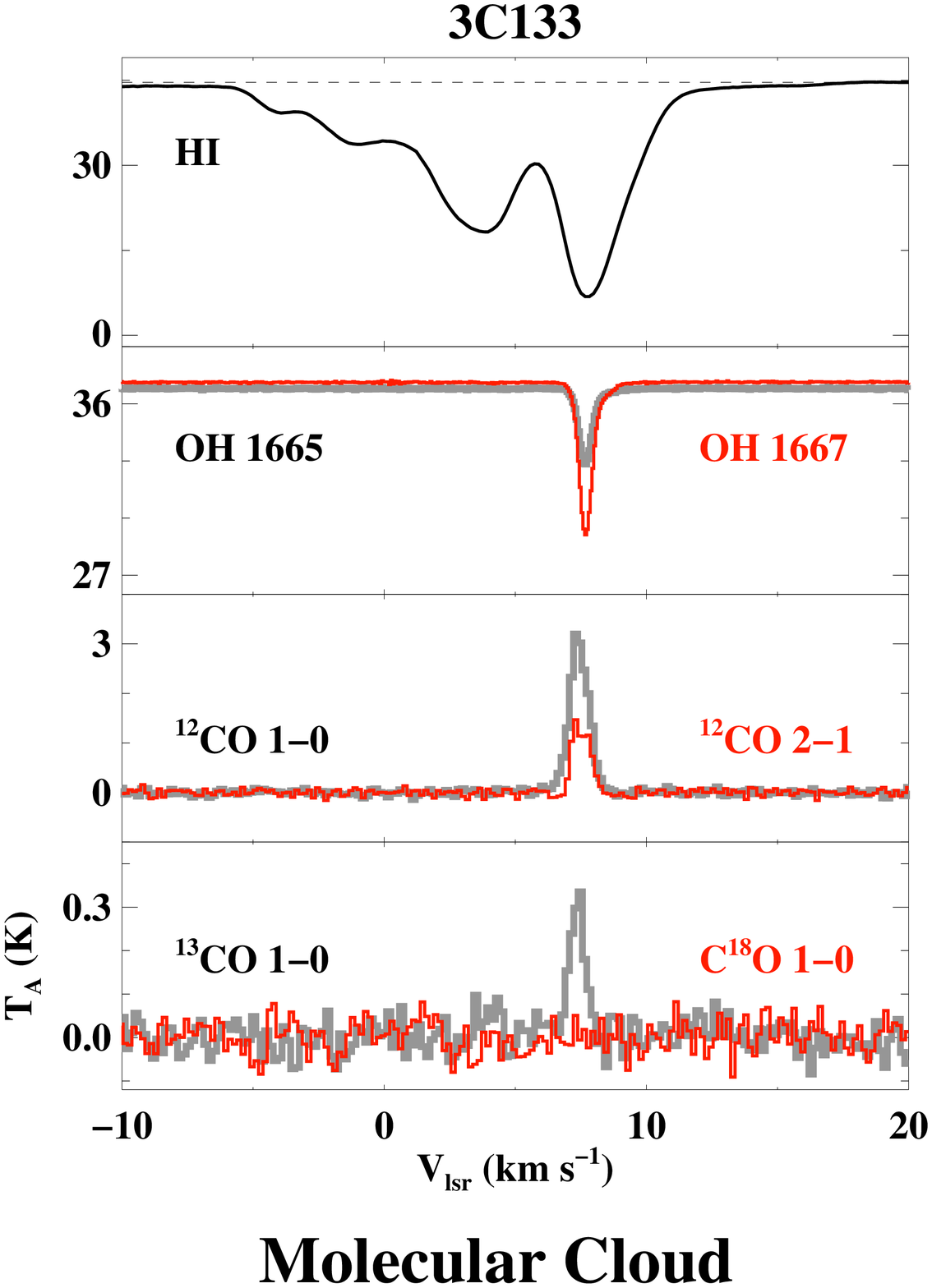}
\caption{Representative spectra as described in section \ref{subsec:darkgas}. The 3C192 sightline has only \hi\ seen in absorption. One component of the 3C133 sightline has OH and \hi\ in absorption and CO and its isotopologues in emission. The 3C132 sightline has one gas component with both OH and \hi\ , but no detectable CO transitions.}
\label{fig4}
\end{figure*}
We compare the Gaussian components seen in \hi\ absorption, OH absorption, and CO emission.
A total of 219  \hi\ and 49  
OH absorption components were detected. Most OH components can be associated with an HI component within a velocity offset of ~1.0 \kms, except for three sources with offsets of about 1.5 \kms. There are four general categories of clouds, three of which are illustrated in Fig.~\ref{fig4}. Toward 3C192, only \hi\ is present, typical of CNM. 
Toward 3C133, \hi, OH, and several CO and CO isotopologues are all detected, which is
representative of `normal' molecular clouds. Toward 3C132, there exists a component with \hi\ and OH absorption, but no CO emission, representing DMG. An example of a DMG-threshold cloud can be found in the spectra of 3C109 (Fig.\ \ref{fig:spec1}), where only weak (~0.1 K) CO is detected. In our sample, there are 77.6\% CNM,  4.1\% DMG, 6.9\% DMG-threshold and 11.4\% molecular clouds. In terms of detectability in absorption, the apparent DMG clouds (DMG and DMG-threshold) are similar to molecular clouds with CO emission. 

\begin{figure}
\includegraphics[width=1.0\linewidth]{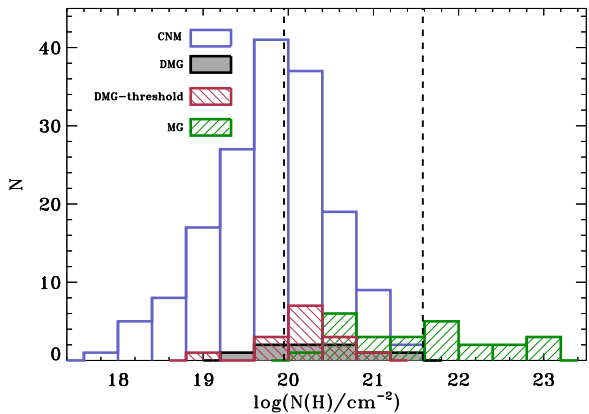}
\caption{Histogram of hydrogen column density, $N$(H), for CNM (blue), DMG (filled gray), DMG-threshold (red), and molecular cloud (MG, green). $N$(H) contains contribution from \hi\ for CNM, DMG, and DMG-threshold cloud. Both \hi\ and H$_2$ (transforming from CO measurements, N(H$_2$)=N(CO)/$1\times 10^{-4}$) are included for MG cloud. The left and right dashed lines represent N(H)= 9.4$\times 10^{19}$ cm$^{-2}$ (visual extinction A$\rm_V=0.05$ mag) and 3.8$\times 10^{21}$ cm$^{-2}$ (A$\rm_V=2$ mag), respectively. } 
\label{fig:hist_hi_dg_mo}
\end{figure}

The statistics of hydrogen column density are presented in Figure \ref{fig:hist_hi_dg_mo}. 
The hydrogen column density of DMG clouds (DMG and DMG-threshold) falls between N(H) $\sim$10$^{20}$ \cm2\ and $4\times10^{21}$ \cm2\, corresponding to a extinction range A$_V \sim$ 0.05 to 2 mag. Within this ``intermediate'' extinction range, the self-shielding of H$_2$ is complete, while that of CO is not. All $N$(H) in this work refer to hydrogen column density based on \hi\ absorption and CO emission measurements. Comparison with $N$(H$_2$) based on Planck results will be published in a separate paper (Hiep et al. in prep). 
The abundance of measured CO is thus expected to be less than the canonical value of 10$^{-4}$ and to vary with extinction. Our detection statistics of OH in DMGs are consistent with  a picture in which gas in this intermediate extinction range is still evolving chemically. 
OH is thus a potentially good tracer of diffuse gas with intermediate extinction, namely, between the self-shielding thresholds of \h2\ and CO. Such a suggestion has been borne out by detailed studies of individual regions with more complete information. \citet{Xu2016}, for example, found much tighter X-factors in both OH and CH than in CO, for gases in the intermediate extinction region of the Taurus molecular cloud.

\section{Implication of Absorption Survey}
\label{sec:abs_survey}

The expected locations and abundance of OH should make it an excellent tracer of DMG. However, due both to low optical depths and low contrast between the main line excitation temperatures and the Galactic diffuse continuum background, OH is generally difficult to detect in emission. Large-scale mapping of OH therefore requires extremely high sensitivity observations and/or the existence of bright continuum background against which the lines may be seen in absorption. A handful of studies  have accomplished the former over very limited regions in the Northern Hemisphere local ISM \citep{Barriault2010,Allen2012,Allen2015,Cotten2012}. The SPLASH project \citep{Dawson2014} has greatly improved on the sensitivities of previous large-scale surveys in the south, to map OH (primarily in absorption) over $> 150$ square degrees of the bright inner Galaxy. However, for outer-Galaxy and off-plane regions, the most practical approach 
will make use of upcoming radio telescopes to conduct comprehensive absorption surveys, of the kind piloted here. 

\begin{figure*}
\includegraphics[width=1.0\linewidth]{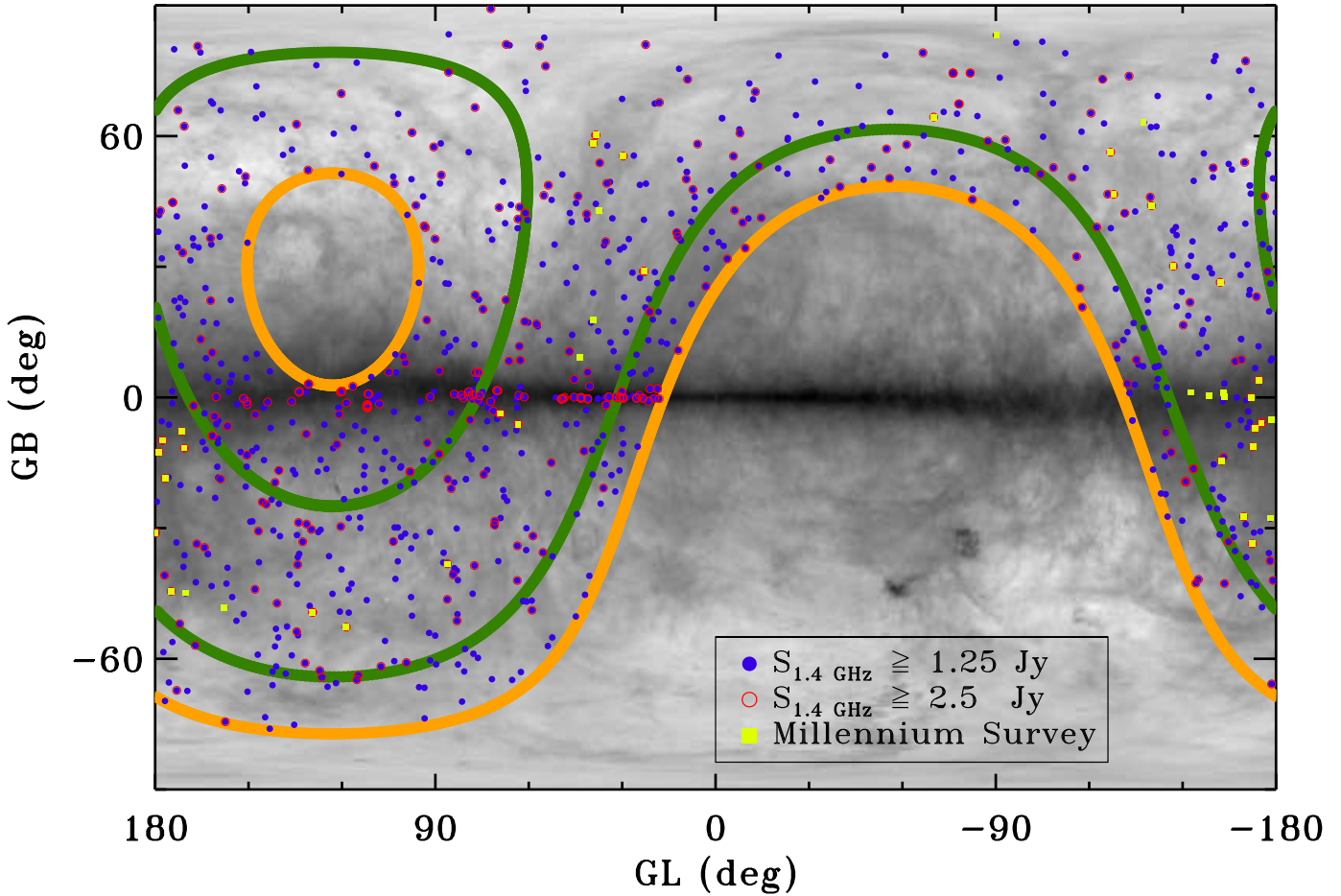}
\caption{Distribution of continuum point sources within the area of FAST sky coverage (limits shown with orange ), which covers a declination range of -14.35 to 65.65 degrees. Red circles represent 372 sources with flux densities greater than 2.5 Jy in the NVSS survey. In initial observation periods, FAST will adopt a drift scan mode. The threshold of 2.5 Jy corresponds to a 3$\sigma$ detection in OH absorption for gas with an optical depth of 0.01, in a drift scan of 12 s, at a velocity resolution of 0.25 km s$^{-1}$, with a system temperature of 25 K. Blue filled circles represent 1071 sources with flux densities greater than 1.25 Jy in the NVSS survey. The threshold of 1.25 Jy allows for a 3$\sigma$ detection for OH  having optical depth of 5.5$\times 10^{-3}$ in a total observing time of 10 minutes (ON+OFF)  in tracking mode. The grey background is the integrated \hi\ intensity map from the LAB \hi\ survey \citep{Hartmann1997,Arnal2000,Bajaja2005}.  The limits of the coverage of Arecibo are shown with  green solid lines.  The positions of the 44 point sources used in this paper are plotted with yellow filled squares. }\vspace{0.5cm}
\label{fig:fast_survey}
\end{figure*}

\begin{figure}
\centering
\includegraphics[width=0.95\linewidth]{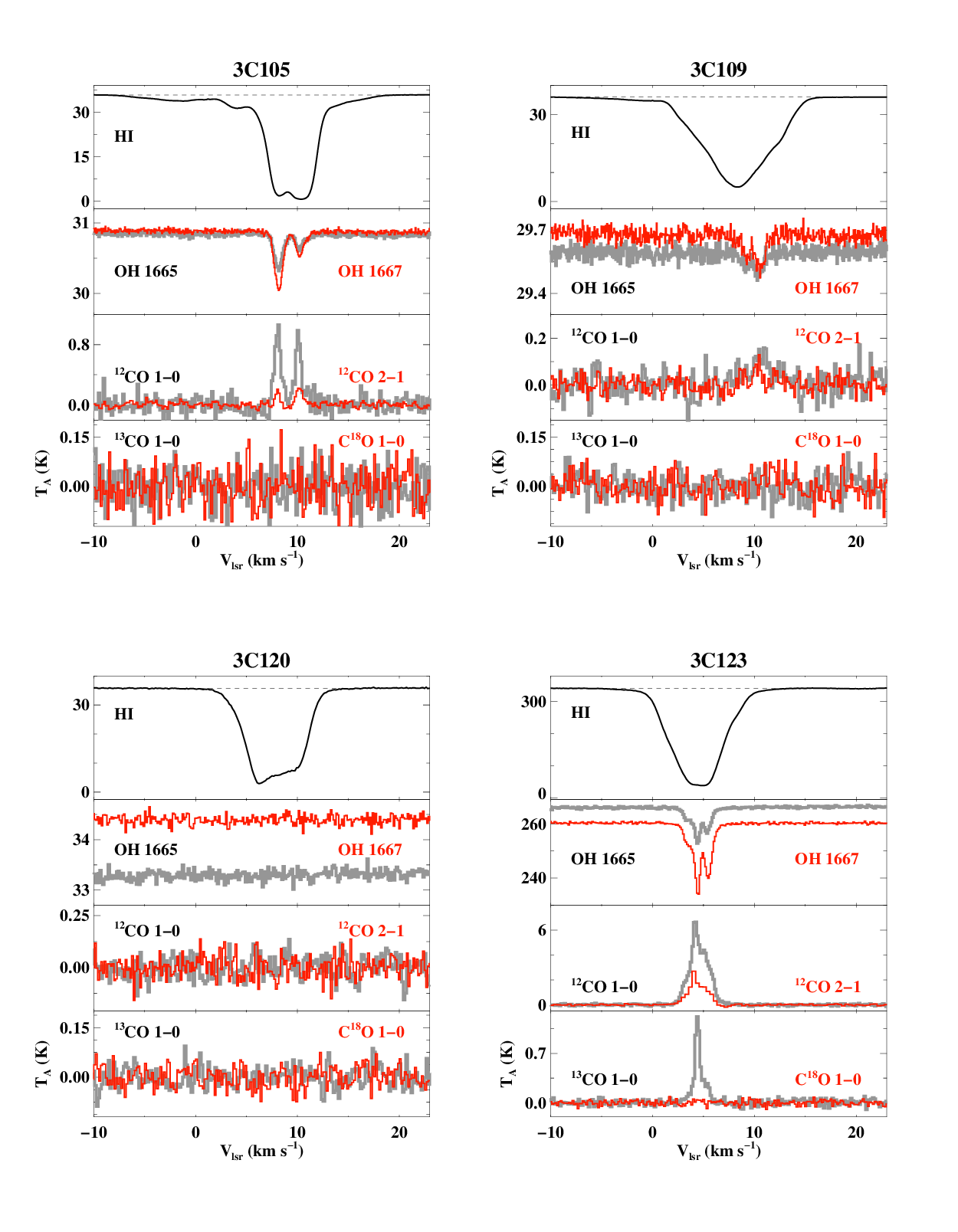}
\caption{Spectra toward 3C105 ($Top\ left$), 3C109 ($(Top\ right)$), 3C120 ($Bottom\ left$) and 3C123 ($Bottom\  right$).The Y axis, $T\rm_A$, represents antenna temperature, but  for CO(1-0) spectra  is the main-beam brightness temperature,  which has been corrected for the main-beam efficiency.}
\label{fig:spec1} 
\end{figure}

The Five-hundred-meter Aperture Spherical radio Telescope (FAST)
commenced observing in September, 2016. The unprecedented sensitivity of FAST and its early science instruments \citep{Li2013} should make feasible an \hi+OH absorption survey, in the mode of the Millennium Survey, but with 10 times more sources. Figure \ref{fig:fast_survey} shows the distribution of potential continuum sources available to FAST.  In the coming decades, the SKA1 will provide the survey speed and sensitivity to measure absorption with a source density of between a few to a few tens per square degree \citep{McClure-Griffiths2015}. This makes feasible an all-sky ``absorption-image", mapping out a fine grid of ISM   excitation temperature and  column density over a very large fraction of the sky. Based on similar excitation and sensitivity considerations, ALMA is a powerful instrument to obtain systematic and sensitive absorption measurements of millimeter lines in diffuse gas. CO and HCO$^+$ in diffuse gas, in particular, will be much better constrained in terms of excitation temperature and column densities through ALMA absorption observations than through emission measurements. Combining both radio and millimeter absorption surveys in the coming decade, we will quantify the DMG and provide definitive answers to questions like the global star formation efficiency. 

\section{Conclusions}
\label{sec:conclusion}

Utilizing unpublished OH absorption measurements from the Millennium Survey and our own follow-up CO surveys, we carried out an analysis of the excitation conditions and quantity of OH  along 44  
sightlines through the Local ISM and Galactic Plane. CO was observed towards  these positions. 49 
OH components were detected towards 22 
of these sightlines. The conclusions are as follows: 

\begin{enumerate}
\item  The excitation temperature of OH peaks around 3.4 K  and follows  a modified 
normalized log-normal distribution, 
\begin{equation}
f(T_\mathrm{ex}) \propto \frac{1}{ \sqrt{2\pi}\sigma} \rm{exp}\left[-\frac{[ln(T_{ex})-ln(3.4 K)]^2}{2\sigma^2}\right]. \nonumber
\end{equation}
The majority of OH gas in our sample,  presumably representative of the Milky Way, thus has an excitation temperature close to the background (CMB plus synchrotron), providing an explanation of why OH has historically been so difficult to detect in emission.
\item The OH main lines are generally not in LTE, with a moderate excitation temperature difference of $|T\rm_{ex}(1667)-T\rm_{ex}(1665)| < 2$ K.
\item The OH emission is optically thin;  the distribution of $\tau_{1667}$ peaks at $\sim0.01$, with the highest value in our sample equal to 0.22. 
\item A weak correlation between $N$(OH) and $N$(\hi) was found. The abundance ratio [OH]/[HI] has a median of $10^{-7}$.

\item $N$(OH) and $N$(CO) are linearly correlated when both are detected, which is consistent with previous observations.
\item Whether a cloud is designated as DMG depends on the sensitivity of the CO data. By comparing with the CfA CO survey data of \citet{Dame2001} we find that the fraction of DMG components would increase by a factor of $\sim$2.5 compared to our results had this less sensitive dataset been used. To highlight this difference, we designated clouds that would be ``CO-dark'' in the CfA survey as ``DMG threshold'' clouds in this work.

\item About 49\% of all detected OH absorption clouds are CO-dark, namely, either DMG or DMG-threshold. The absence of CO emission toward these OH components implies that OH serves as a more effective tracer than CO of diffuse molecular gas within A$_V \sim$ 0.05 to 2 mag.

\item Given the low opacity and the low excitation temperature of the Galactic OH gas, sensitive absorption surveys made feasible by upcoming large telescopes, such as FAST and SKA, are needed for a comprehensive inventory of Galactic cold gas.

\end{enumerate}

\section*{Acknowledgments}

This work is supported by National Key R\&D Program of China 2017YFA0402600 and International Partnership Program of Chinese Academy of Sciences  No.\ 114A11KYSB20160008. D.L.\ acknowledges support from "CAS Interdisciplinary Innovation Team'' program. J.R.D.\ is the
recipient of an Australian Research Council DECRA Fellowship
(project number DE170101086). This work was carried out in part at the Jet Propulsion Laboratory, which is operated for NASA by the California Institute of Technology. L.B.\  acknowledges support from CONICYT Project PFB06 
CO data were observed with the Delingha 13.7m telescope of the Qinghai Station of Purple Mountain Observatory (PMODLH), the Caltech Submillimeter Observatory (CSO), and the IRAM 30-meter telescope. The authors appreciate all the staff members of the PMODLH, CSO, and the IRAM 30-meter Observatory for their help during the observations. We thank Lei Qian and Lei Zhu for their help in CSO observations.

\clearpage

\begin{figure}
\centering
\includegraphics[width=0.98\linewidth]{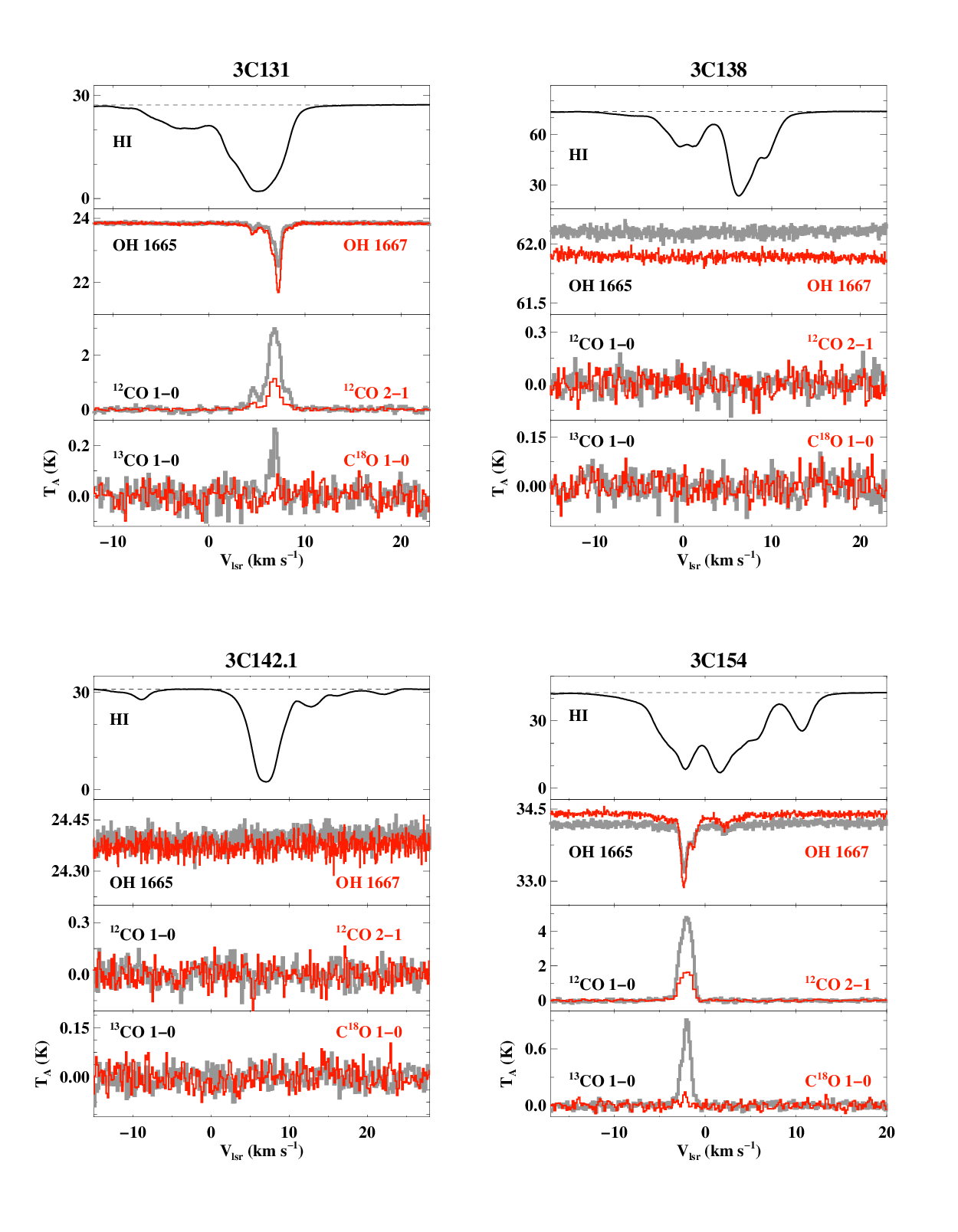}
\caption{Spectra toward 3C131 ($Top\ left$), 3C138 ($(Top\ right)$), 3C142.1 ($Bottom\ left$) and 3C154 ($Bottom\ right$). The Y axis, $T\rm_A$, represents antenna temperature, but  for CO(1-0) spectra  is the main-beam brightness temperature,  which has been corrected for the main-beam efficiency.}
\label{fig:spec2} 
\end{figure}

\begin{figure}
\centering
\includegraphics[width=0.98\linewidth]{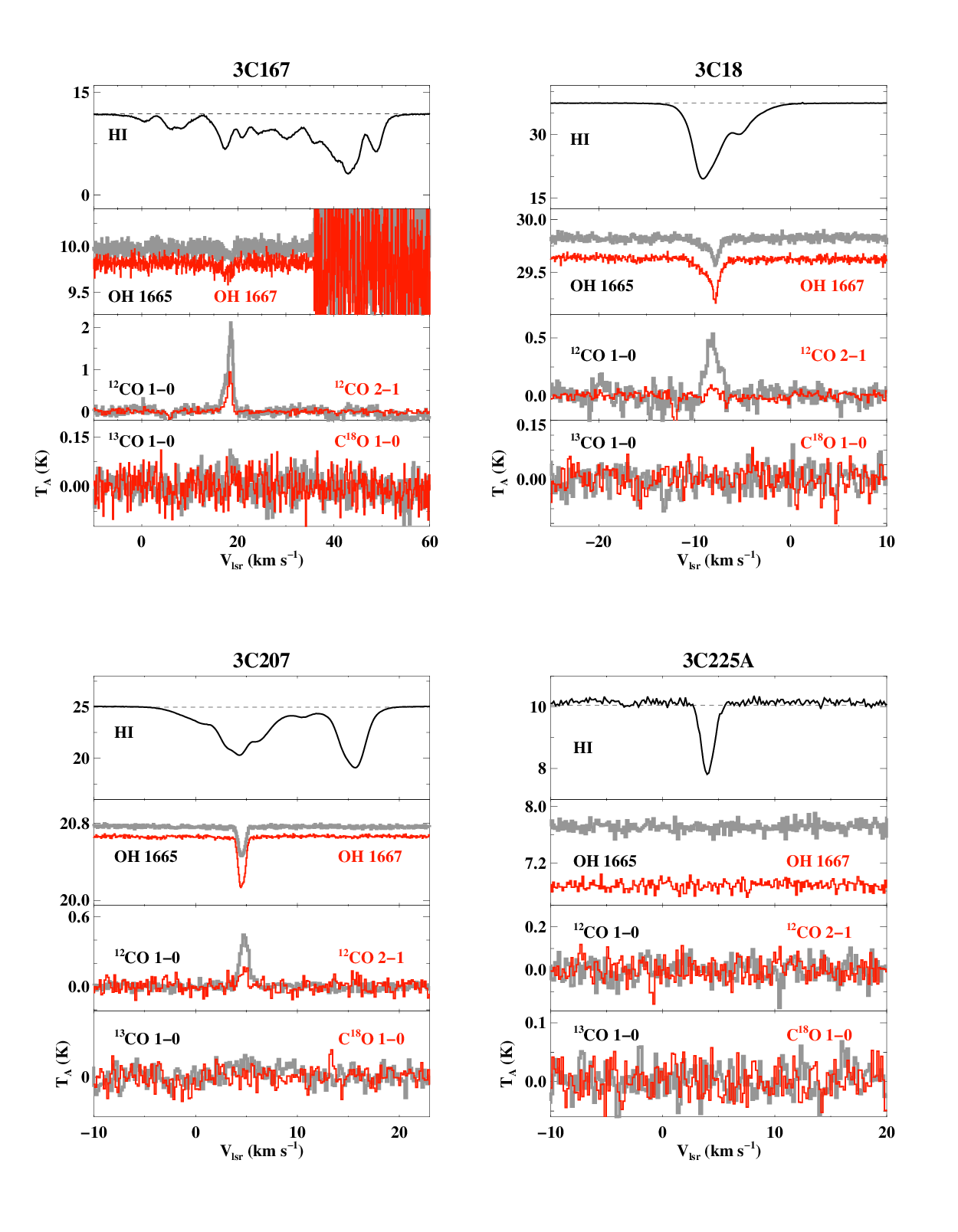}
\caption{Spectra toward 3C167 ($Top\ left$), 3C18 ($(Top\ right)$), 3C207 ($Bottom\ left$) and 3C225A ($Bottom\ right$). The Y axis, $T\rm_A$, represents antenna temperature, but  for CO(1-0) spectra  is the main-beam brightness temperature,  which has been corrected for the main-beam efficiency.}
\label{fig:spec3} 
\end{figure}

\begin{figure}
\centering
\includegraphics[width=0.98\linewidth]{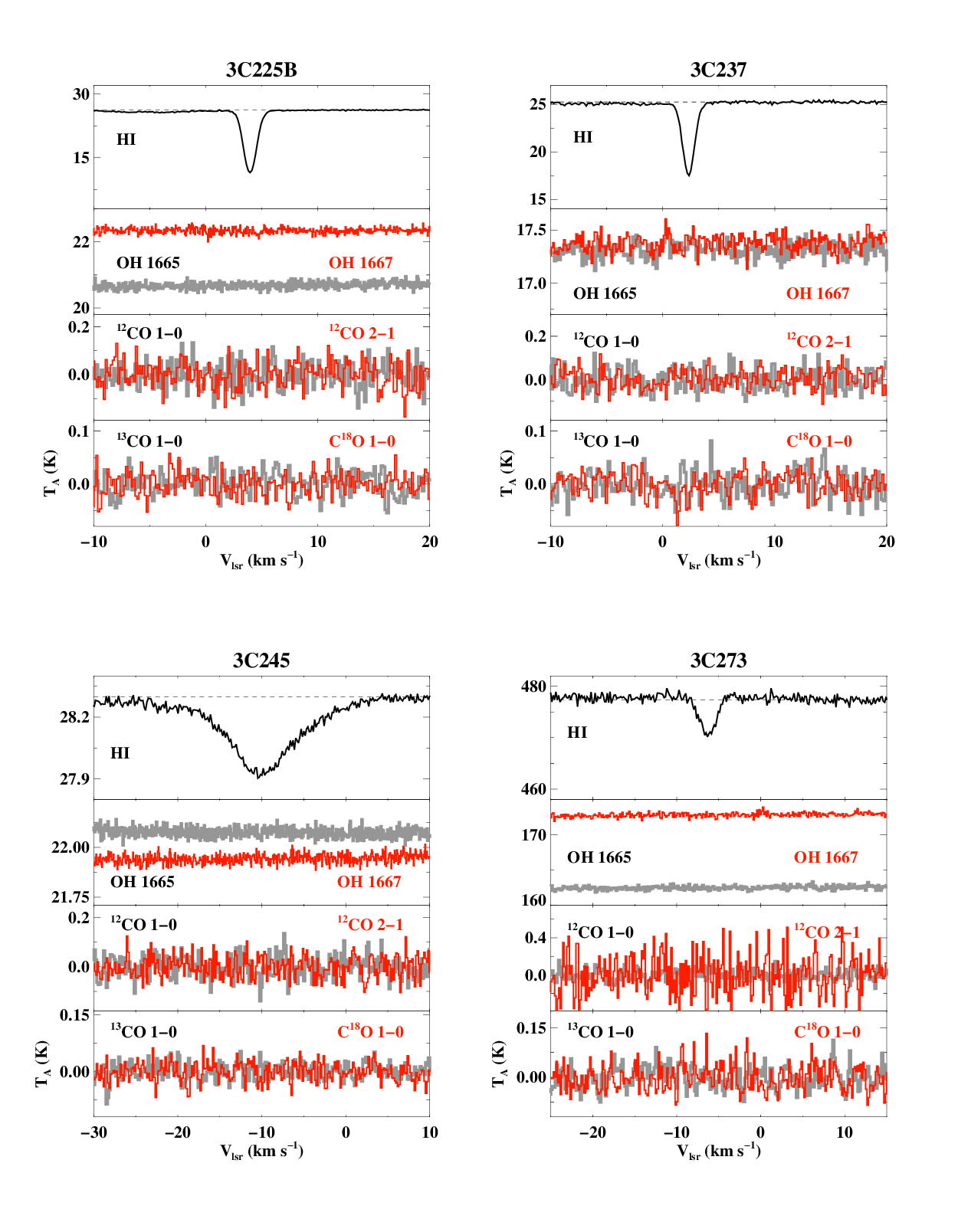}
\caption{Spectra toward 3C225B ($Top\ left$), 3C237 ($(Top\ right)$), 3C245 ($Bottom\ left$) and 3C273 ($Bottom\ right$). The Y axis, $T\rm_A$, represents antenna temperature, but  for CO(1-0) spectra  is the main-beam brightness temperature,  which has been corrected for the main-beam efficiency.}
\label{fig:spec4} 
\end{figure}

\begin{figure}
\centering
\includegraphics[width=0.98\linewidth]{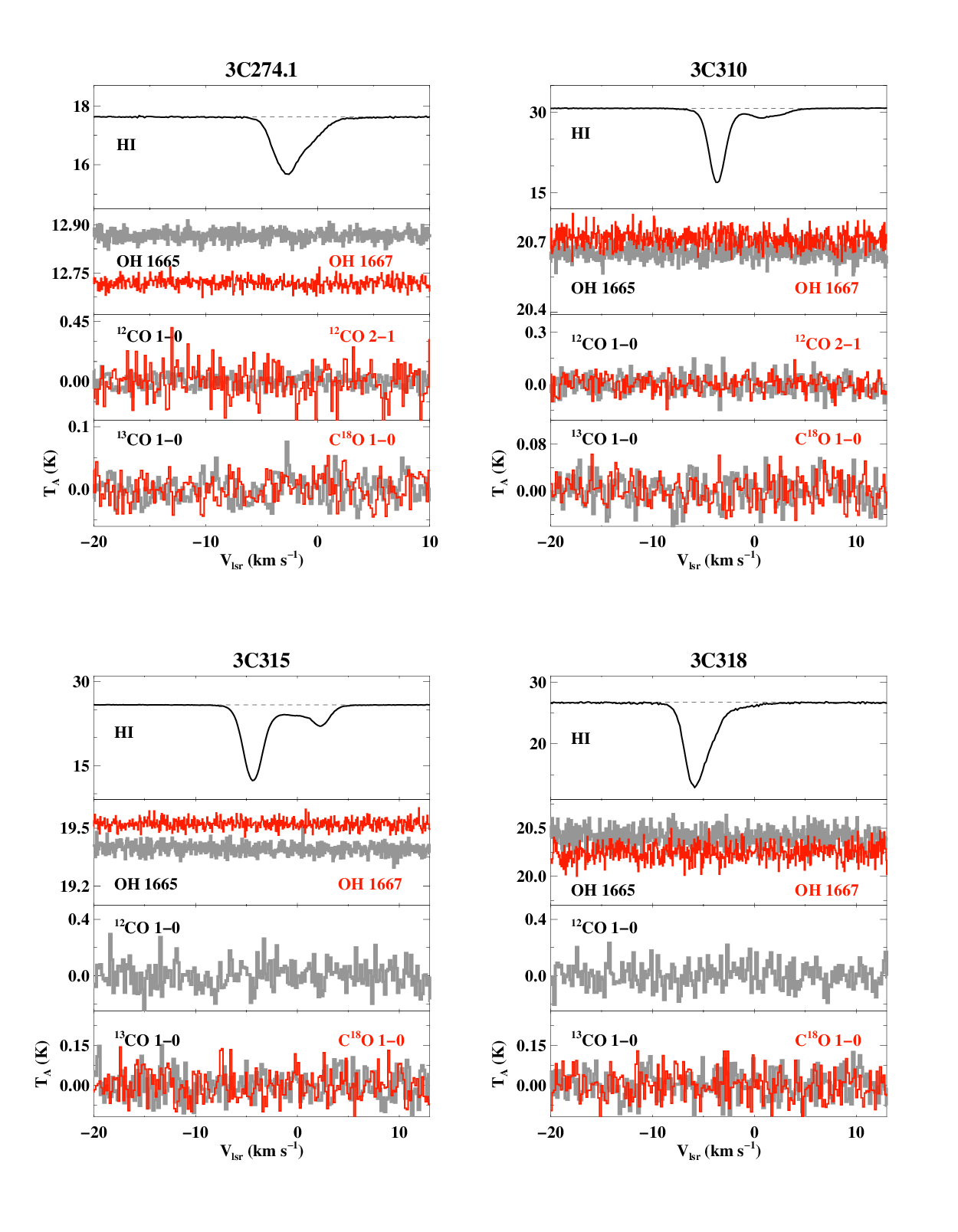}
\caption{Spectra toward 3C274.1 ($Top\ left$), 3C109 ($(Top\ right)$), 3C315 ($Bottom\ left$) and 3C318 ($Bottom\ right$). The Y axis, $T\rm_A$, represents antenna temperature, but  for CO(1-0) spectra  is the main-beam brightness temperature,  which has been corrected for the main-beam efficiency.}
\label{fig:spec5} 
\end{figure}

\begin{figure}
\centering
\includegraphics[width=0.98\linewidth]{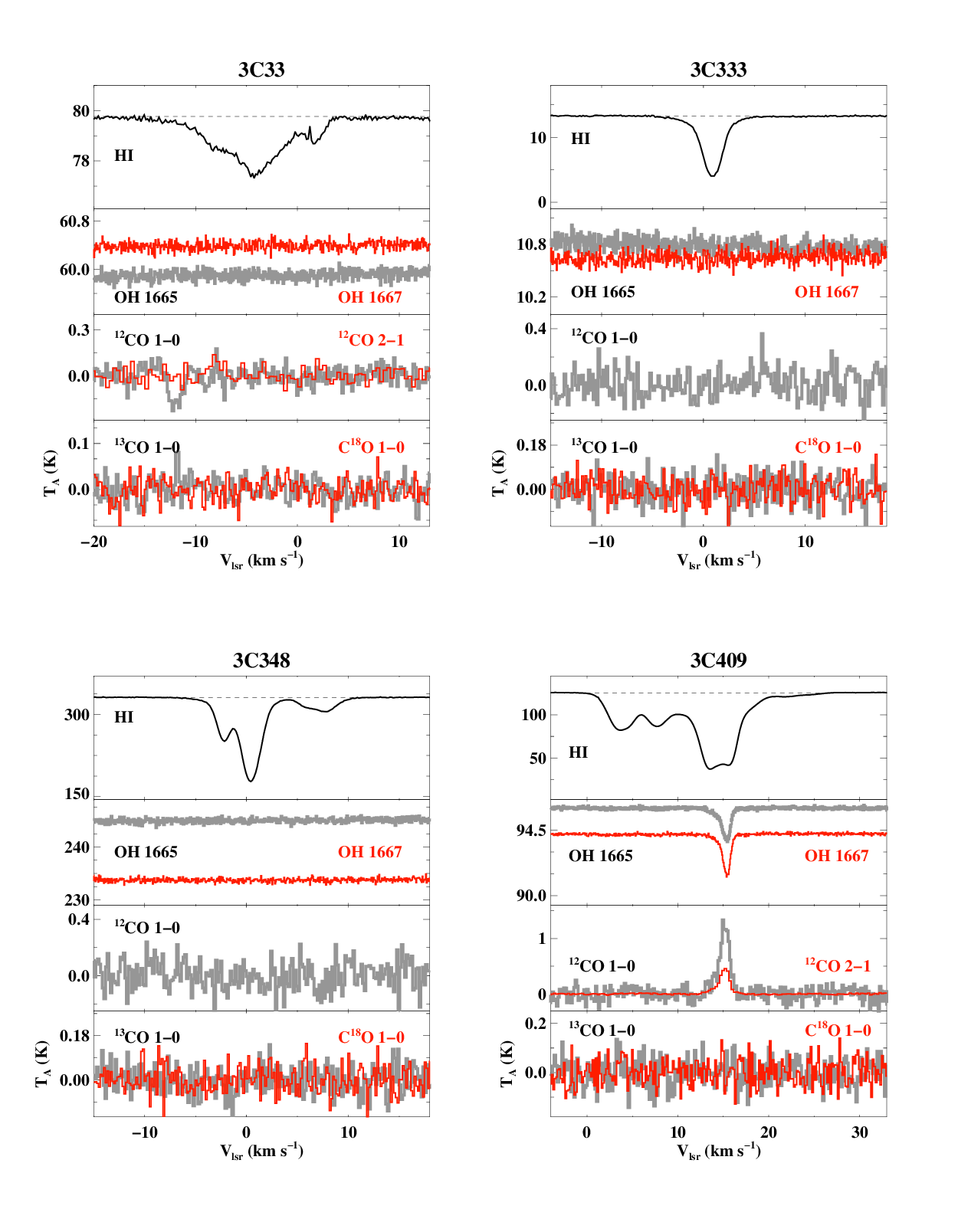}
\caption{Spectra toward 3C33 ($Top\ left$), 3C333 ($(Top\ right)$), 3C348 ($Bottom\ left$) and 3C409 ($Bottom\ right$). The Y axis, $T\rm_A$, represents antenna temperature, but  for CO(1-0) spectra  is the main-beam brightness temperature,  which has been corrected for the main-beam efficiency.}
\label{fig:spec6} 
\end{figure}

\begin{figure}
\centering
\includegraphics[width=0.98\linewidth]{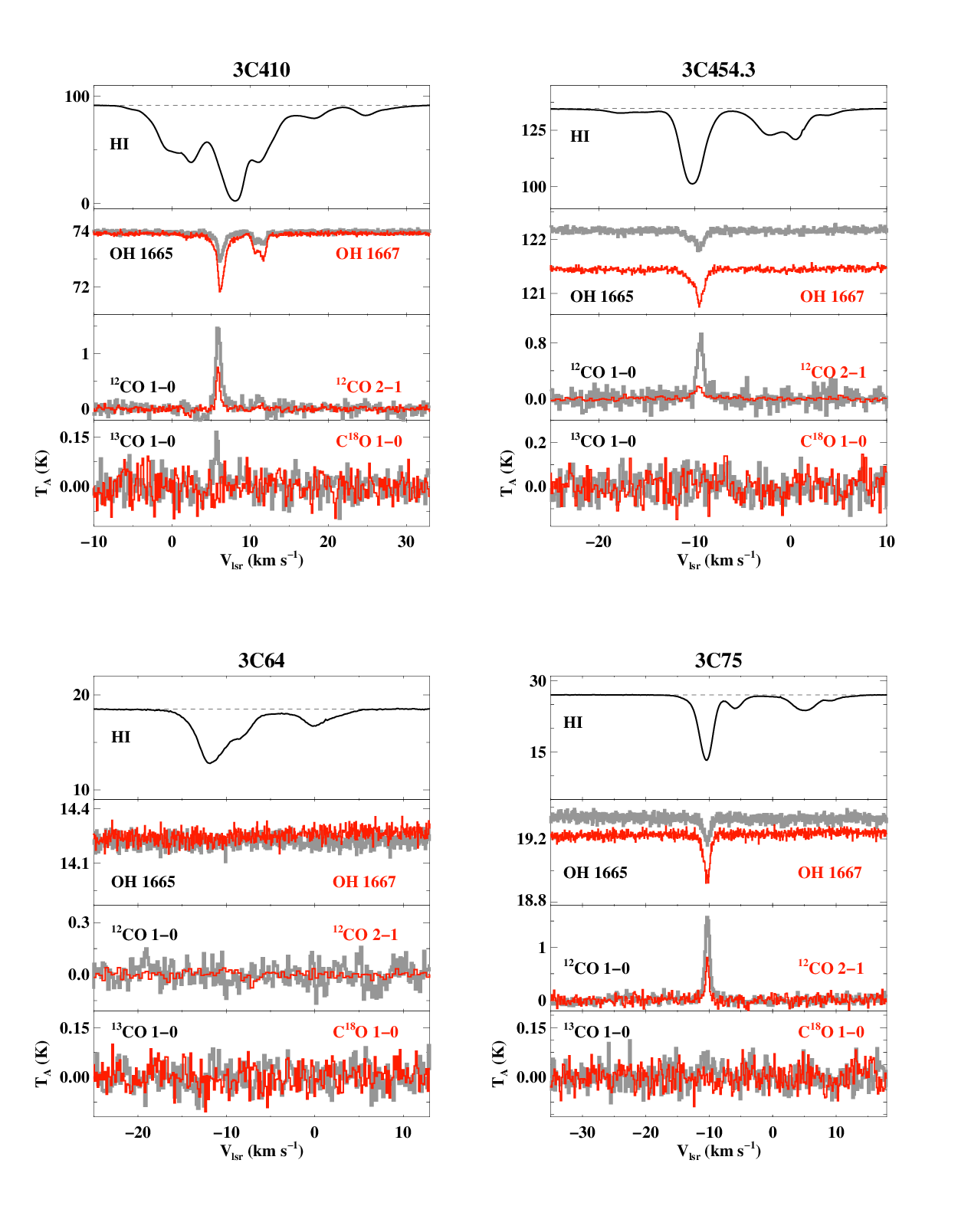}
\caption{Spectra toward 3C410 ($Top\ left$), 3C454.3 ($(Top\ right)$), 3C64 ($Bottom\ left$) and 3C75 ($Bottom\ right$). The Y axis, $T\rm_A$, represents antenna temperature, but  for CO(1-0) spectra  is the main-beam brightness temperature,  which has been corrected for the main-beam efficiency.}
\label{fig:spec7} 
\end{figure}

\begin{figure}
\centering
\includegraphics[width=0.98\linewidth]{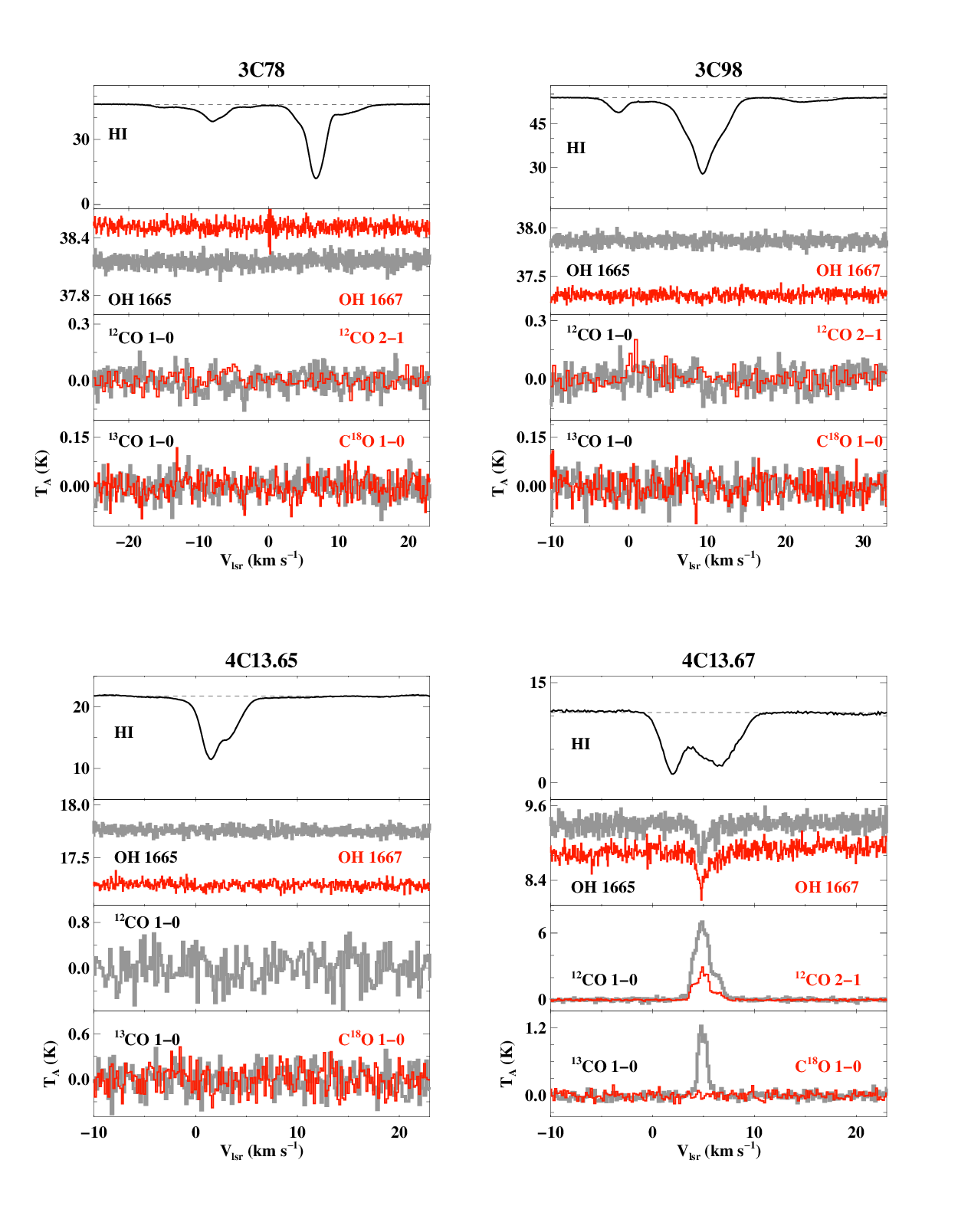}
\caption{Spectra toward 3C78 ($Top\ left$), 3C98 ($(Top\ right)$), 4C13.65 ($Bottom\ left$) and 4C13.67 ($Bottom\ right$). The Y axis, $T\rm_A$, represents antenna temperature, but  for CO(1-0) spectra  is the main-beam brightness temperature,  which has been corrected for the main-beam efficiency.}
\label{fig:spec8} 
\end{figure}

\begin{figure}
\centering
\includegraphics[width=0.98\linewidth]{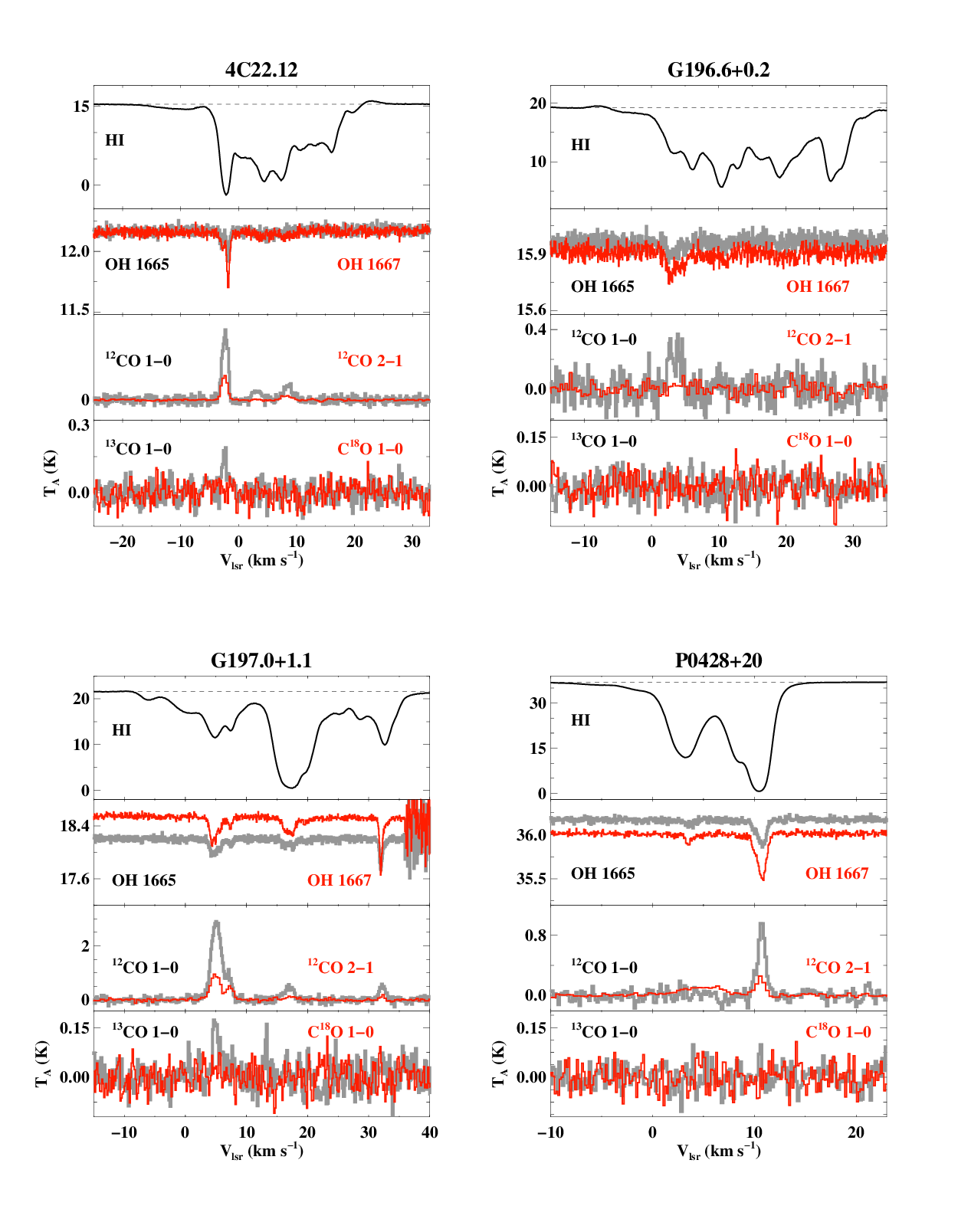}
\caption{Spectra toward 4C22.12 ($Top\ left$), G196.6+0.2 ($(Top\ right)$), G197.0+1.1 ($Bottom\ left$) and P0428+20 ($Bottom\ right$). The Y axis, $T\rm_A$, represents antenna temperature, but  for CO(1-0) spectra  is the main-beam brightness temperature,  which has been corrected for the main-beam efficiency.}
\label{fig:spec9} 
\end{figure}

\begin{figure}
\centering
\includegraphics[width=0.98\linewidth]{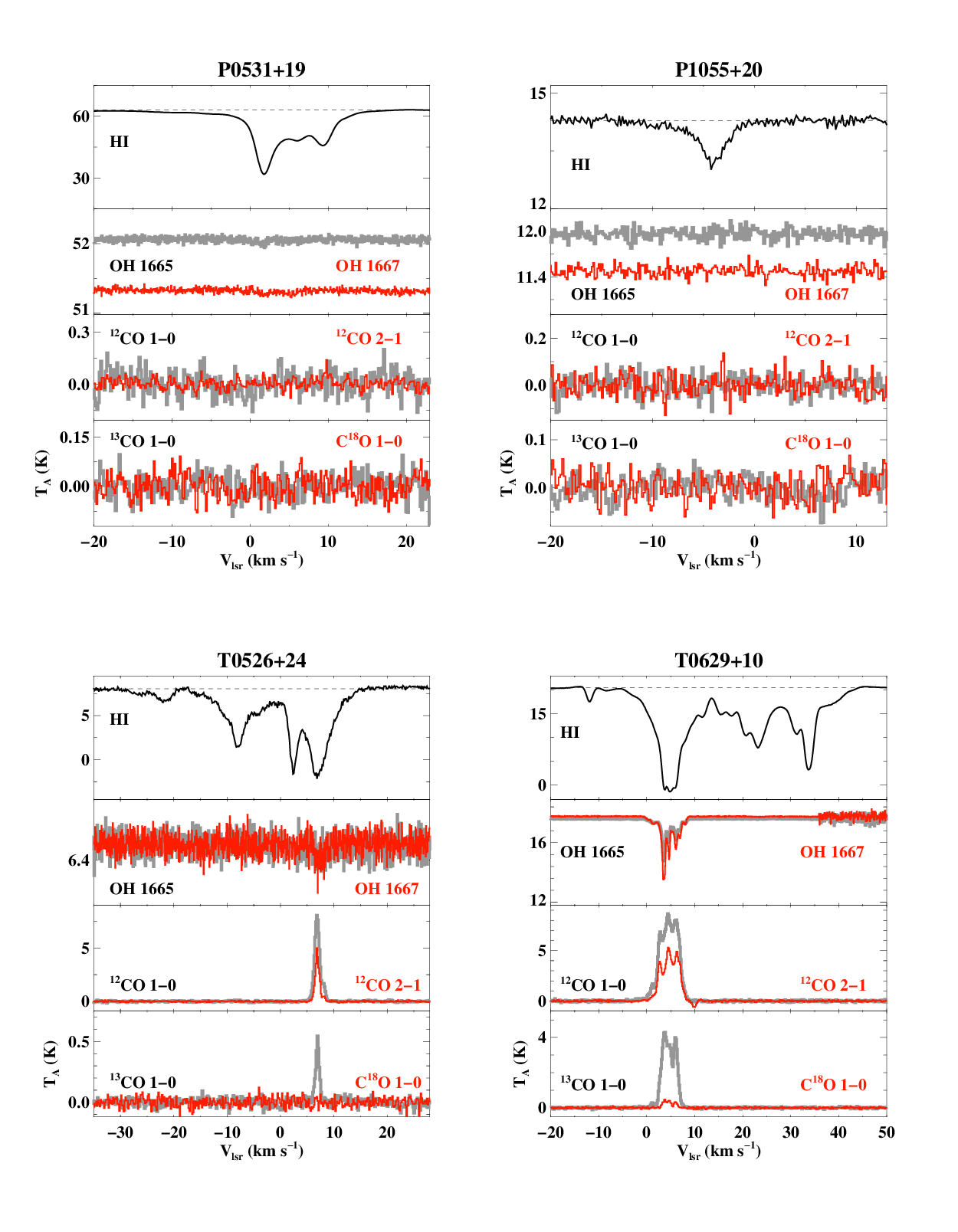}
\caption{Spectra toward P0531+19 ($Top\ left$), P1055+20 ($(Top\ right)$), T0526+24 ($Bottom\ left$) and T0629+10 ($Bottom\ right$).The Y axis, $T\rm_A$, represents antenna temperature, but  for CO(1-0) spectra  is the main-beam brightness temperature,  which has been corrected for the main-beam efficiency.}
\label{fig:spec10} 
\end{figure}

\begin{figure}
\centering
\includegraphics[width=0.49\linewidth]{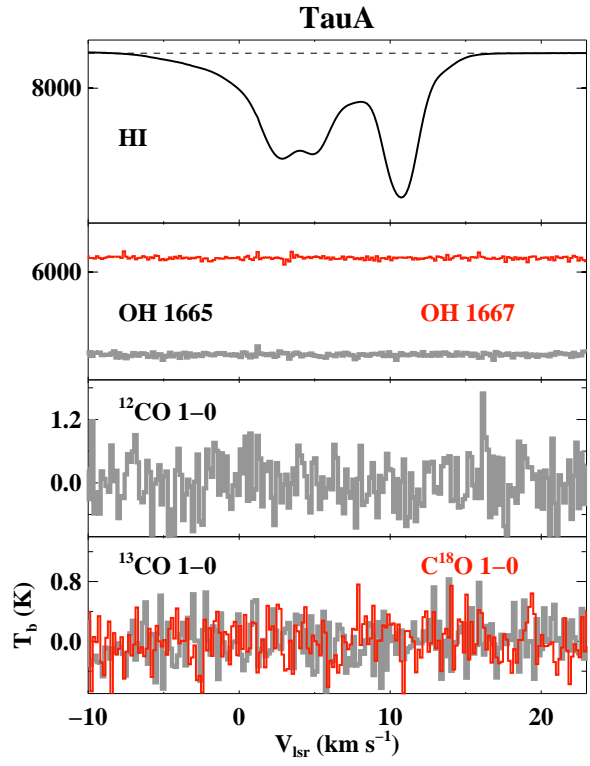}
\caption{Spectra toward TauA. The Y axis, $T\rm_A$, represents antenna temperature, but  for CO(1-0) spectra  is the main-beam brightness temperature,  which has been corrected for the main-beam efficiency.}
\label{fig:spec11} 
\end{figure}

\begin{sidewaystable}[h!]
\fontsize{7}{6}\selectfont
\centering
\caption{Gaussian Fit Parameters for OH main lines}
\label{table:1}
\renewcommand{\arraystretch}{1.5}
\begin{tabular}{ lllllllclllll  }
 \hline
 \hline
\multirow{2}{*}{$Source$} & \multicolumn{1}{c}{\multirow{2}{*}{$l/b$}} & \multicolumn{6}{c}{OH(1665)} & \multicolumn{5}{c}{OH(1667)}\\
\cline{3-7}
\cline{9-13}
 & & \multicolumn{1}{c}{$\tau$} & \multicolumn{1}{c}{$V_{lsr}$} & \multicolumn{1}{c}{$\Delta V$} & \multicolumn{1}{c}{$T_{ex}$} & \multicolumn{1}{c}{$N(OH)$} & & \multicolumn{1}{c}{$\tau$} & \multicolumn{1}{c}{$V_{lsr}$} & \multicolumn{1}{c}{$\Delta V$} & \multicolumn{1}{c}{$T_{ex}$} & \multicolumn{1}{c}{$N(OH)$} \\
 
(Name) & \multicolumn{1}{c}{$(^{o})$} & \multicolumn{1}{c}{$\ $} & \multicolumn{1}{c}{$(km s^{-1})$} & \multicolumn{1}{c}{$(km s^{-1})$} & \multicolumn{1}{c}{$(K)$} & \multicolumn{1}{c}{$(10^{14}cm^{-2})$} & & \multicolumn{1}{c}{$\ $} & \multicolumn{1}{c}{$(km s^{-1})$} & \multicolumn{1}{c}{$(km s^{-1})$} & \multicolumn{1}{c}{$(K)$} & \multicolumn{1}{c}{$(10^{14}cm^{-2})$} \\ 
\hline

3C105 & 187.6/-33.6 & 0.0158 $\pm$ 0.0003 & 8.14 $\pm$ 0.01 & 0.97 $\pm$ 0.02 & 4.93 $\pm$ 0.5 & 0.32 $\pm$ 0.26 & &  0.0268 $\pm$ 0.0003 & 8.17 $\pm$ 0.0 & 0.96 $\pm$ 0.01 & 3.96 $\pm$ 0.34 & 0.24 $\pm$ 0.13 \\
3C105 & 187.6/-33.6 & 0.0062 $\pm$ 0.0003 & 10.23 $\pm$ 0.02 & 1.03 $\pm$ 0.05 & 8.49 $\pm$ 1.26 & 0.23 $\pm$ 0.44 & &  0.0107 $\pm$ 0.0003 & 10.26 $\pm$ 0.01 & 1.03 $\pm$ 0.03 & 7.64 $\pm$ 0.88 & 0.2 $\pm$ 0.22 \\
3C109 & 181.8/-27.8 & 0.0022 $\pm$ 0.0003 & 9.16 $\pm$ 0.1 & 1.0 $\pm$ 0.24 & 23.68 $\pm$ 3.66 & 0.22 $\pm$ 0.73 & &  0.0036 $\pm$ 0.0004 & 9.22 $\pm$ 0.07 & 1.02 $\pm$ 0.17 & 24.13 $\pm$ 2.5 & 0.21 $\pm$ 0.36 \\
3C109 & 181.8/-27.8 & 0.0035 $\pm$ 0.0003 & 10.44 $\pm$ 0.06 & 0.95 $\pm$ 0.15 & 14.76 $\pm$ 2.33 & 0.21 $\pm$ 0.56 & &  0.0057 $\pm$ 0.0004 & 10.58 $\pm$ 0.05 & 1.04 $\pm$ 0.11 & 13.92 $\pm$ 1.58 & 0.19 $\pm$ 0.29 \\
3C123 & 170.6/-11.7 & 0.0191 $\pm$ 0.0008 & 3.65 $\pm$ 0.06 & 1.19 $\pm$ 0.11 & 11.08 $\pm$ 1.79 & 1.08 $\pm$ 1.26 & &  0.0347 $\pm$ 0.0012 & 3.71 $\pm$ 0.06 & 1.22 $\pm$ 0.1 & 11.2 $\pm$ 1.26 & 1.13 $\pm$ 0.69 \\
3C123 & 170.6/-11.7 & 0.0431 $\pm$ 0.0023 & 4.43 $\pm$ 0.01 & 0.53 $\pm$ 0.03 & 6.42 $\pm$ 1.22 & 0.62 $\pm$ 0.57 & &  0.0919 $\pm$ 0.004 & 4.46 $\pm$ 0.01 & 0.53 $\pm$ 0.02 & 7.31 $\pm$ 0.75 & 0.84 $\pm$ 0.29 \\
3C123 & 170.6/-11.7 & 0.0338 $\pm$ 0.0008 & 5.37 $\pm$ 0.01 & 0.92 $\pm$ 0.04 & 11.62 $\pm$ 1.11 & 1.53 $\pm$ 0.8 & &  0.0784 $\pm$ 0.0012 & 5.47 $\pm$ 0.01 & 0.92 $\pm$ 0.02 & 8.13 $\pm$ 0.61 & 1.38 $\pm$ 0.37 \\
3C131 & 171.4/-7.8 & 0.0065 $\pm$ 0.0005 & 4.55 $\pm$ 0.02 & 0.56 $\pm$ 0.06 & 12.67 $\pm$ 1.53 & 0.2 $\pm$ 0.3 & &  0.0103 $\pm$ 0.0006 & 4.56 $\pm$ 0.02 & 0.62 $\pm$ 0.05 & 7.51 $\pm$ 1.04 & 0.11 $\pm$ 0.16 \\
3C131 & 171.4/-7.8 & 0.0074 $\pm$ 0.0006 & 6.8 $\pm$ 0.06 & 2.97 $\pm$ 0.2 & 10.46 $\pm$ 0.86 & 0.98 $\pm$ 0.94 & &  0.0111 $\pm$ 0.0005 & 6.68 $\pm$ 0.05 & 3.32 $\pm$ 0.15 & 11.08 $\pm$ 0.59 & 0.96 $\pm$ 0.49 \\
3C131 & 171.4/-7.8 & 0.0167 $\pm$ 0.0007 & 6.59 $\pm$ 0.01 & 0.42 $\pm$ 0.02 & 2.66 $\pm$ 0.84 & 0.08 $\pm$ 0.19 & &  0.0259 $\pm$ 0.0008 & 6.58 $\pm$ 0.01 & 0.43 $\pm$ 0.02 & 4.15 $\pm$ 0.58 & 0.11 $\pm$ 0.1 \\
3C131 & 171.4/-7.8 & 0.0521 $\pm$ 0.0007 & 7.23 $\pm$ 0.0 & 0.55 $\pm$ 0.01 & 5.03 $\pm$ 0.25 & 0.62 $\pm$ 0.13 & &  0.0858 $\pm$ 0.0007 & 7.23 $\pm$ 0.0 & 0.59 $\pm$ 0.01 & 5.43 $\pm$ 0.16 & 0.65 $\pm$ 0.07 \\
3C132 & 178.9/-12.5 & 0.0033 $\pm$ 0.0002 & 7.82 $\pm$ 0.03 & 0.89 $\pm$ 0.08 & 15.72 $\pm$ 2.08 & 0.19 $\pm$ 0.45 & &  0.0056 $\pm$ 0.0003 & 7.79 $\pm$ 0.02 & 0.81 $\pm$ 0.05 & 23.1 $\pm$ 1.27 & 0.25 $\pm$ 0.18 \\
3C133 & 177.7/-9.9 & 0.1009 $\pm$ 0.001 & 7.66 $\pm$ 0.0 & 0.53 $\pm$ 0.0 & 2.97 $\pm$ 0.22 & 0.68 $\pm$ 0.16 & &  0.2132 $\pm$ 0.0016 & 7.68 $\pm$ 0.0 & 0.52 $\pm$ 0.0 & 1.33 $\pm$ 1.24 & 0.35 $\pm$ 0.7 \\
3C133 & 177.7/-9.9 & 0.0148 $\pm$ 0.001 & 7.94 $\pm$ 0.02 & 1.23 $\pm$ 0.03 & 5.9 $\pm$ 3.41 & 0.46 $\pm$ 2.18 & &  0.0333 $\pm$ 0.0015 & 7.96 $\pm$ 0.02 & 1.23 $\pm$ 0.02 & 6.11 $\pm$ 2.22 & 0.59 $\pm$ 1.18 \\
3C154 & 185.6/4.0 & 0.0266 $\pm$ 0.0006 & -2.32 $\pm$ 0.01 & 0.74 $\pm$ 0.03 & 2.61 $\pm$ 0.57 & 0.22 $\pm$ 0.3 & &  0.0428 $\pm$ 0.0007 & -2.34 $\pm$ 0.01 & 0.71 $\pm$ 0.02 & 2.69 $\pm$ 0.35 & 0.19 $\pm$ 0.12 \\
3C154 & 185.6/4.0 & 0.01 $\pm$ 0.0005 & -1.39 $\pm$ 0.04 & 0.83 $\pm$ 0.09 & 6.15 $\pm$ 1.43 & 0.22 $\pm$ 0.51 & &  0.0179 $\pm$ 0.0006 & -1.34 $\pm$ 0.03 & 0.91 $\pm$ 0.06 & 4.7 $\pm$ 0.74 & 0.18 $\pm$ 0.21 \\
3C154 & 185.6/4.0 & 0.0038 $\pm$ 0.0004 & 2.23 $\pm$ 0.06 & 1.12 $\pm$ 0.16 & 6.62 $\pm$ 3.22 & 0.12 $\pm$ 0.95 & &  0.0054 $\pm$ 0.0005 & 2.18 $\pm$ 0.06 & 1.33 $\pm$ 0.14 & 2.1 $\pm$ 1.99 & 0.04 $\pm$ 0.46 \\
3C167 & 207.3/1.2 & 0.0109 $\pm$ 0.0019 & 18.53 $\pm$ 0.11 & 1.27 $\pm$ 0.26 & 4.87 $\pm$ 2.23 & 0.29 $\pm$ 1.26 & &  0.0108 $\pm$ 0.0013 & 17.66 $\pm$ 0.16 & 2.71 $\pm$ 0.4 & 4.65 $\pm$ 1.49 & 0.32 $\pm$ 1.0 \\
3C18 & 118.6/-52.7 & 0.003 $\pm$ 0.0003 & -8.53 $\pm$ 0.11 & 2.5 $\pm$ 0.19 & 10.93 $\pm$ 1.99 & 0.35 $\pm$ 1.16 & &  0.006 $\pm$ 0.0003 & -8.33 $\pm$ 0.05 & 2.5 $\pm$ 0.1 & 9.27 $\pm$ 0.86 & 0.33 $\pm$ 0.39 \\
3C18 & 118.6/-52.7 & 0.0056 $\pm$ 0.0004 & -7.82 $\pm$ 0.02 & 0.68 $\pm$ 0.06 & 2.05 $\pm$ 2.03 & 0.03 $\pm$ 0.44 & &  0.0078 $\pm$ 0.0005 & -7.85 $\pm$ 0.01 & 0.6 $\pm$ 0.04 & 1.0 $\pm$ 0.0 & 0.01 $\pm$ 0.0 \\
3C207 & 213.0/30.1 & 0.0152 $\pm$ 0.0002 & 4.55 $\pm$ 0.01 & 0.77 $\pm$ 0.01 & 3.17 $\pm$ 0.42 & 0.16 $\pm$ 0.17 & &  0.0268 $\pm$ 0.0002 & 4.55 $\pm$ 0.0 & 0.78 $\pm$ 0.01 & 2.57 $\pm$ 0.2 & 0.13 $\pm$ 0.06 \\
3C409 & 63.4/-6.1 & 0.0062 $\pm$ 0.0007 & 14.63 $\pm$ 0.14 & 1.86 $\pm$ 0.15 & 11.37 $\pm$ 1.9 & 0.55 $\pm$ 1.19 & &  0.0057 $\pm$ 0.0008 & 14.7 $\pm$ 0.18 & 1.68 $\pm$ 0.18 & 7.8 $\pm$ 3.86 & 0.18 $\pm$ 1.15 \\
3C409 & 63.4/-6.1 & 0.0199 $\pm$ 0.0012 & 15.4 $\pm$ 0.01 & 0.89 $\pm$ 0.04 & 0.17 $\pm$ 0.86 & 0.01 $\pm$ 0.46 & &  0.0273 $\pm$ 0.0016 & 15.41 $\pm$ 0.01 & 0.86 $\pm$ 0.03 & 0.2 $\pm$ 0.0 & 0.01 $\pm$ 0.0 \\
3C410 & 69.2/-3.8 & 0.0121 $\pm$ 0.0003 & 6.25 $\pm$ 0.01 & 1.03 $\pm$ 0.03 & 9.4 $\pm$ 0.97 & 0.5 $\pm$ 0.47 & &  0.025 $\pm$ 0.0004 & 6.29 $\pm$ 0.01 & 1.19 $\pm$ 0.02 & 4.31 $\pm$ 0.43 & 0.3 $\pm$ 0.19 \\
3C410 & 69.2/-3.8 & 0.0043 $\pm$ 0.0003 & 10.7 $\pm$ 0.04 & 0.7 $\pm$ 0.09 & 11.26 $\pm$ 3.32 & 0.14 $\pm$ 0.65 & &  0.0084 $\pm$ 0.0005 & 10.72 $\pm$ 0.04 & 0.8 $\pm$ 0.09 & 4.42 $\pm$ 1.55 & 0.07 $\pm$ 0.27 \\
3C410 & 69.2/-3.8 & 0.0053 $\pm$ 0.0003 & 11.67 $\pm$ 0.04 & 0.84 $\pm$ 0.09 & 5.74 $\pm$ 2.44 & 0.11 $\pm$ 0.64 & &  0.0114 $\pm$ 0.0005 & 11.68 $\pm$ 0.03 & 0.81 $\pm$ 0.07 & 2.92 $\pm$ 1.14 & 0.06 $\pm$ 0.23 \\
3C454.3 & 86.1/-38.2 & 0.0023 $\pm$ 0.0001 & -9.67 $\pm$ 0.03 & 1.63 $\pm$ 0.07 & 4.61 $\pm$ 2.14 & 0.07 $\pm$ 0.72 & &  0.0044 $\pm$ 0.0001 & -9.55 $\pm$ 0.02 & 1.37 $\pm$ 0.04 & 8.25 $\pm$ 1.22 & 0.12 $\pm$ 0.26 \\
3C75 & 170.3/-44.9 & 0.0072 $\pm$ 0.0003 & -10.36 $\pm$ 0.03 & 1.32 $\pm$ 0.07 & 3.64 $\pm$ 1.09 & 0.15 $\pm$ 0.52 & &  0.0142 $\pm$ 0.0003 & -10.36 $\pm$ 0.02 & 1.25 $\pm$ 0.04 & 3.91 $\pm$ 0.6 & 0.16 $\pm$ 0.21 \\
4C13.67 & 43.5/9.2 & 0.0474 $\pm$ 0.0037 & 4.88 $\pm$ 0.05 & 1.3 $\pm$ 0.12 & 10.43 $\pm$ 0.9 & 2.73 $\pm$ 1.13 & &  0.057 $\pm$ 0.004 & 4.95 $\pm$ 0.05 & 1.46 $\pm$ 0.12 & 10.37 $\pm$ 0.62 & 2.04 $\pm$ 0.56 \\
4C22.12 & 188.1/0.0 & 0.0059 $\pm$ 0.0009 & -2.84 $\pm$ 0.06 & 0.79 $\pm$ 0.17 & 6.91 $\pm$ 2.35 & 0.14 $\pm$ 0.61 & &  0.0103 $\pm$ 0.001 & -2.73 $\pm$ 0.04 & 0.8 $\pm$ 0.11 & 6.75 $\pm$ 1.28 & 0.13 $\pm$ 0.25 \\
4C22.12 & 188.1/0.0 & 0.0172 $\pm$ 0.0011 & -1.78 $\pm$ 0.02 & 0.56 $\pm$ 0.05 & 4.67 $\pm$ 0.95 & 0.19 $\pm$ 0.3 & &  0.0356 $\pm$ 0.0012 & -1.78 $\pm$ 0.01 & 0.54 $\pm$ 0.02 & 3.8 $\pm$ 0.46 & 0.17 $\pm$ 0.11 \\
G196.6+0.2 & 196.6/0.2 & 0.0043 $\pm$ 0.0005 & 3.22 $\pm$ 0.1 & 1.67 $\pm$ 0.25 & 10.95 $\pm$ 2.34 & 0.34 $\pm$ 1.1 & &  0.0065 $\pm$ 0.0005 & 3.43 $\pm$ 0.09 & 2.5 $\pm$ 0.24 & 8.85 $\pm$ 1.24 & 0.34 $\pm$ 0.59 \\
G197.0+1.1 & 197.0/1.1 & 0.0124 $\pm$ 0.0005 & 4.83 $\pm$ 0.03 & 1.84 $\pm$ 0.09 & 5.2 $\pm$ 0.65 & 0.5 $\pm$ 0.57 & &  0.0188 $\pm$ 0.0007 & 4.72 $\pm$ 0.03 & 1.59 $\pm$ 0.07 & 5.51 $\pm$ 0.5 & 0.39 $\pm$ 0.26 \\
G197.0+1.1 & 197.0/1.1 & 0.0057 $\pm$ 0.0008 & 7.46 $\pm$ 0.04 & 0.61 $\pm$ 0.1 & 1.0 $\pm$ 0.0 & 0.01 $\pm$ 0.0 & &  0.0075 $\pm$ 0.0011 & 7.34 $\pm$ 0.04 & 0.6 $\pm$ 0.1 & 1.0 $\pm$ 0.0 & 0.01 $\pm$ 0.0 \\
G197.0+1.1 & 197.0/1.1 & 0.0053 $\pm$ 0.0006 & 16.41 $\pm$ 0.09 & 1.17 $\pm$ 0.22 & 8.76 $\pm$ 1.91 & 0.23 $\pm$ 0.69 & &  0.0094 $\pm$ 0.0017 & 16.34 $\pm$ 0.12 & 0.95 $\pm$ 0.19 & 7.07 $\pm$ 1.33 & 0.15 $\pm$ 0.29 \\
G197.0+1.1 & 197.0/1.1 & 0.0069 $\pm$ 0.0008 & 17.65 $\pm$ 0.05 & 0.71 $\pm$ 0.12 & 5.35 $\pm$ 1.88 & 0.11 $\pm$ 0.47 & &  0.0124 $\pm$ 0.001 & 17.45 $\pm$ 0.12 & 1.22 $\pm$ 0.21 & 7.15 $\pm$ 0.89 & 0.26 $\pm$ 0.29 \\
G197.0+1.1 & 197.0/1.1 & 0.0237 $\pm$ 0.0009 & 32.01 $\pm$ 0.01 & 0.56 $\pm$ 0.02 & 3.41 $\pm$ 0.62 & 0.19 $\pm$ 0.23 & &  0.0428 $\pm$ 0.0012 & 32.01 $\pm$ 0.01 & 0.53 $\pm$ 0.02 & 4.35 $\pm$ 0.39 & 0.23 $\pm$ 0.1 \\
P0428+20 & 176.8/-18.6 & 0.0015 $\pm$ 0.0002 & 3.6 $\pm$ 0.08 & 1.06 $\pm$ 0.2 & 13.3 $\pm$ 4.09 & 0.09 $\pm$ 0.71 & &  0.0029 $\pm$ 0.0003 & 3.55 $\pm$ 0.04 & 0.75 $\pm$ 0.09 & 4.56 $\pm$ 2.56 & 0.02 $\pm$ 0.25 \\
P0428+20 & 176.8/-18.6 & 0.0076 $\pm$ 0.0002 & 10.7 $\pm$ 0.02 & 1.1 $\pm$ 0.04 & 13.29 $\pm$ 0.79 & 0.47 $\pm$ 0.32 & &  0.0137 $\pm$ 0.0003 & 10.7 $\pm$ 0.01 & 1.11 $\pm$ 0.02 & 11.71 $\pm$ 0.45 & 0.42 $\pm$ 0.14 \\
P0531+19 & 186.8/-7.1 & 0.0011 $\pm$ 0.00024 & 1.559 $\pm$ 0.087 & 0.814 $\pm$ 0.214 & 3.49 $\pm$ 4.37 & 0.0133 $\pm$ 0.0169 & &  0.00089 $\pm$ 0.00024 & 1.85 $\pm$ 0.139 & 1.053 $\pm$ 0.346 & 4.01 $\pm$ 7.1 & 0.0089 $\pm$ 0.0162 \\
T0526+24 & 181.4/-5.2 & 0.0158 $\pm$ 0.0058 & 7.58 $\pm$ 0.29 & 1.66 $\pm$ 0.74 & 13.57 $\pm$ 2.78 & 1.51 $\pm$ 2.61 & &  0.0359 $\pm$ 0.0059 & 7.49 $\pm$ 0.15 & 1.97 $\pm$ 0.39 & 10.21 $\pm$ 1.2 & 1.7 $\pm$ 1.15 \\
T0629+10 & 201.5/0.5 & 0.004 $\pm$ 0.0026 & 0.16 $\pm$ 0.19 & 0.64 $\pm$ 0.49 & 4.07 $\pm$ 2.23 & 0.05 $\pm$ 0.39 & &  0.0113 $\pm$ 0.0023 & 0.33 $\pm$ 0.22 & 0.97 $\pm$ 0.43 & 3.22 $\pm$ 0.96 & 0.08 $\pm$ 0.24 \\
T0629+10 & 201.5/0.5 & 0.0378 $\pm$ 0.0129 & 3.13 $\pm$ 0.28 & 0.99 $\pm$ 0.43 & 0.59 $\pm$ 0.66 & 0.09 $\pm$ 0.54 & &  0.0769 $\pm$ 0.0046 & 4.17 $\pm$ 0.06 & 3.17 $\pm$ 0.22 & 0.92 $\pm$ 1.87 & 0.53 $\pm$ 3.88 \\
T0629+10 & 201.5/0.5 & 0.0168 $\pm$ 0.0016 & 1.47 $\pm$ 0.11 & 1.43 $\pm$ 0.44 & 1.49 $\pm$ 0.46 & 0.15 $\pm$ 0.37 & &  0.0206 $\pm$ 0.0034 & 1.3 $\pm$ 0.1 & 0.8 $\pm$ 0.2 & 0.5 $\pm$ 0.0 & 0.02 $\pm$ 0.01 \\
T0629+10 & 201.5/0.5 & 0.163 $\pm$ 0.0281 & 3.61 $\pm$ 0.02 & 0.61 $\pm$ 0.04 & 2.33 $\pm$ 0.13 & 0.98 $\pm$ 0.22 & &  0.2147 $\pm$ 0.0045 & 3.55 $\pm$ 0.0 & 0.6 $\pm$ 0.02 & 3.1 $\pm$ 0.13 & 0.95 $\pm$ 0.09 \\
T0629+10 & 201.5/0.5 & 0.0812 $\pm$ 0.002 & 4.62 $\pm$ 0.01 & 0.78 $\pm$ 0.03 & 1.23 $\pm$ 0.29 & 0.33 $\pm$ 0.27 & &  0.1004 $\pm$ 0.0044 & 4.64 $\pm$ 0.01 & 0.36 $\pm$ 0.02 & 1.48 $\pm$ 0.26 & 0.13 $\pm$ 0.07 \\
T0629+10 & 201.5/0.5 & 0.075 $\pm$ 0.0018 & 6.09 $\pm$ 0.02 & 1.05 $\pm$ 0.05 & 3.74 $\pm$ 0.2 & 1.25 $\pm$ 0.25 & &  0.1011 $\pm$ 0.0043 & 6.1 $\pm$ 0.01 & 0.64 $\pm$ 0.04 & 4.46 $\pm$ 0.25 & 0.69 $\pm$ 0.13 \\
T0629+10 & 201.5/0.5 & 0.0371 $\pm$ 0.0031 & 6.99 $\pm$ 0.02 & 0.49 $\pm$ 0.06 & 4.28 $\pm$ 0.29 & 0.33 $\pm$ 0.13 & &  0.0744 $\pm$ 0.0036 & 6.93 $\pm$ 0.02 & 0.64 $\pm$ 0.05 & 3.99 $\pm$ 0.22 & 0.45 $\pm$ 0.1 \\
T0629+10 & 201.5/0.5 & 0.0174 $\pm$ 0.0018 & 7.9 $\pm$ 0.05 & 0.83 $\pm$ 0.13 & 3.54 $\pm$ 0.46 & 0.22 $\pm$ 0.22 & &  0.0295 $\pm$ 0.0025 & 7.95 $\pm$ 0.03 & 0.71 $\pm$ 0.08 & 3.46 $\pm$ 0.42 & 0.17 $\pm$ 0.12 \\

\hline
\end{tabular}
\end{sidewaystable}


\clearpage 
\begin{sidewaystable}[h!]
\fontsize{5}{4}\selectfont
\centering
\caption{ Gaussian Fit Parameters of CO Data}
\label{table:2}
\renewcommand{\arraystretch}{1.5}
\begin{tabular}{ lllllllccllllcc  }
 \hline
 \hline
\multirow{2}{*}{$Source$} & \multicolumn{1}{c}{\multirow{2}{*}{$l/b$}} & \multicolumn{4}{c}{$^{12}$CO(1-0)} &  \multicolumn{4}{c}{$^{12}$CO(2-1)}   & \multicolumn{3}{c}{$^{13}$CO(1-0)} & \multicolumn{1}{c}{\multirow{2}{*}{ $O$ \tablenotemark{b}}} & \multicolumn{1}{c}{\multirow{2}{*}{N($^{12}$CO)}} \\
\cline{3-5}
\cline{7-9}
\cline{11-13}
 & & \multicolumn{1}{c}{$T\rm_{peak}$\tablenotemark{a}} & \multicolumn{1}{c}{$V_{lsr}$} & \multicolumn{1}{c}{$\Delta V$} &  & \multicolumn{1}{c}{$T\rm_{peak}$\tablenotemark{a}} & \multicolumn{1}{c}{$V\rm_{lsr}$} & \multicolumn{1}{c}{$\Delta V$} &  & \multicolumn{1}{c}{$T\rm_{peak}$\tablenotemark{a}} & \multicolumn{1}{c}{$V\rm_{lsr}$} & \multicolumn{1}{c}{$\Delta V$} & & \\

(Name) & \multicolumn{1}{c}{$(^{\circ})$} & \multicolumn{1}{c}{$\rm K $} & \multicolumn{1}{c}{$(\rm km\ s^{-1})$} & \multicolumn{1}{c}{$(\rm km s^{-1})$} & & \multicolumn{1}{c}{$\rm K $} & \multicolumn{1}{c}{$(\rm km\ s^{-1})$} & \multicolumn{1}{c}{$(\rm km\ s^{-1})$} & & \multicolumn{1}{c}{$K $} & \multicolumn{1}{c}{$(\rm km\ s^{-1})$} & \multicolumn{1}{c}{$(\rm km\ s^{-1})$} & & \multicolumn{1}{c}{$(10^{17})\rm\ cm^{-2}$} \\ 
\hline

3C105 & 187.6/-33.6 & 1.0 & 8.07 $\pm$ 0.03 & 0.77 $\pm$ 0.06 &  & 0.3 & 8.05 $\pm$ 0.03 & 0.45 $\pm$ 0.06 &  &  -  &  -  &  -  & 0 & 0.1032 $\pm$  0.0083 \\
3C105 & 187.6/-33.6 & 0.9 & 10.81 $\pm$ 0.03 & 0.68 $\pm$ 0.09 &  & 0.3 & 10.20 $\pm$ 0.03 & 0.74 $\pm$ 0.08 &  &  -  &  -  &  -  & 1 & 0.0651 $\pm$  0.0082 \\
3C109 & 181.8/-27.8 &  -  &  -  &  -  &  &  -  &  -  &  -  &  &  -  &  -  &  -  & 0 &  -  \\
3C109 & 181.8/-27.8 & 0.1 & 10.69 $\pm$ 0.15 & 1.52 $\pm$ 0.42 &  &  -  &  -  &  -  &  &  -  &  -  &  -  & 1 & 0.0206 $\pm$  0.0057 \\
3C123 & 170.6/-11.7 & 2.8 & 3.99 $\pm$ 0.00 & 1.87 $\pm$ 0.03 &  & 1.4 & 3.75 $\pm$ 0.03 & 1.71 $\pm$ 0.10 &  & 0.1 & 3.87 $\pm$    0.23 & 1.33 $\pm$ 0.38 & 0 & 1.7736 $\pm$  0.0312 \\
3C123 & 170.6/-11.7 & 3.2 & 4.18 $\pm$ 0.00 & 0.47 $\pm$ 0.01 &  & 2.2 & 4.07 $\pm$ 0.01 & 0.51 $\pm$ 0.03 &  & 1.1 & 4.41 $\pm$    0.01 & 0.48 $\pm$ 0.02 & 1 & 5.1962 $\pm$  0.1415 \\
3C123 & 170.6/-11.7 & 3.2 & 5.21 $\pm$ 0.01 & 1.62 $\pm$ 0.02 &  & 1.8 & 5.14 $\pm$ 0.03 & 1.43 $\pm$ 0.07 &  & 0.3 & 5.17 $\pm$    0.06 & 0.93 $\pm$ 0.14 & 2 & 3.1336 $\pm$  0.0483 \\
3C131 & 171.4/-7.8 & 0.7 & 4.59 $\pm$ 0.16 & 0.71 $\pm$ 0.16 &  & 0.4 & 4.56 $\pm$ 0.05 & 0.84 $\pm$ 0.13 &  &  -  &  -  &  -  & 0 & 0.0506 $\pm$  0.0113 \\
3C131 & 171.4/-7.8 & 1.8 & 6.88 $\pm$ 0.16 & 2.07 $\pm$ 0.16 &  & 0.9 & 6.92 $\pm$ 0.01 & 1.64 $\pm$ 0.10 &  & 0.2 & 6.81 $\pm$    0.05 & 0.96 $\pm$ 0.12 & 1 & 3.2982 $\pm$  0.3194 \\
3C131 & 171.4/-7.8 & 1.2 & 6.64 $\pm$ 0.16 & 0.72 $\pm$ 0.16 &  & 0.8 & 6.64 $\pm$ 0.03 & 0.67 $\pm$ 0.06 &  &  -  &  -  &  -  & 2 & 0.0842 $\pm$  0.0187 \\
3C131 & 171.4/-7.8 & 0.7 & 7.18 $\pm$ 0.16 & 0.48 $\pm$ 0.16 &  & 0.3 & 7.10 $\pm$ 0.04 & 0.41 $\pm$ 0.01 &  &  -  &  -  &  -  & 3 & 0.0335 $\pm$  0.0112 \\
3C132 & 178.9/-12.5 &  -  &  -  &  -  &  &  -  &  -  &  -  &  &  -  &  -  &  -  & 0 &  -  \\
3C133 & 177.7/-9.9 & 3.1 & 7.43 $\pm$ 0.00 & 0.86 $\pm$ 0.01 &  & 2.0 & 7.50 $\pm$ 0.01 & 0.81 $\pm$ 0.02 &  & 0.3 & 7.36 $\pm$    0.03 & 0.70 $\pm$ 0.06 & 0 & 2.0087 $\pm$  0.0248 \\
3C133 & 177.7/-9.9 &  -  &  -  &  -  &  &  -  &  -  &  -  &  &  -  &  -  &  -  & 1 &  -  \\
3C154 & 185.6/4.0 & 3.7 & -2.41 $\pm$ 0.01 & 1.23 $\pm$ 0.02 &  & 2.1 & -2.37 $\pm$ 0.35 & 1.24 $\pm$ 0.35 &  & 0.9 & -2.06 $\pm$    0.01 & 1.13 $\pm$ 0.03 & 0 & 8.6360 $\pm$  0.1185 \\
3C154 & 185.6/4.0 & 3.0 & -1.62 $\pm$ 0.01 & 0.94 $\pm$ 0.01 &  & 1.5 & -1.57 $\pm$ 0.35 & 0.79 $\pm$ 0.35 &  &  -  &  -  &  -  & 1 & 0.2972 $\pm$  0.0047 \\
3C154 & 185.6/4.0 &  -  &  -  &  -  &  &  -  &  -  &  -  &  &  -  &  -  &  -  & 2 &  -  \\
3C167 & 207.3/1.2 & 0.7 & 17.47 $\pm$ 0.31 & 1.59 $\pm$ 0.57 &  & 0.3 & 17.64 $\pm$ 0.25 & 1.62 $\pm$ 0.32 &  &  -  &  -  &  -  & 0 & 0.1098 $\pm$  0.0394 \\
3C18 & 118.6/-52.7 & 0.5 & -8.22 $\pm$ 0.06 & 1.31 $\pm$ 0.14 &  & 0.1 & -8.32 $\pm$ 0.14 & 0.56 $\pm$ 0.32 &  &  -  &  -  &  -  & 0 & 0.1078 $\pm$  0.0117 \\
3C18 & 118.6/-52.7 & 0.2 & -7.10 $\pm$ 0.08 & 0.40 $\pm$ 0.15 &  & 0.1 & -7.74 $\pm$ 0.17 & 0.33 $\pm$ 0.30 &  &  -  &  -  &  -  & 1 & 0.0082 $\pm$  0.0030 \\
3C207 & 213.0/30.1 & 0.4 & 4.79 $\pm$ 0.02 & 1.00 $\pm$ 0.04 &  & 0.2 & 4.74 $\pm$ 0.09 & 0.80 $\pm$ 0.19 &  &  -  &  -  &  -  & 0 & 0.0443 $\pm$  0.0019 \\
3C409 & 63.4/-6.1 & 0.2 & 13.82 $\pm$ 0.30 & 1.25 $\pm$ 0.44 &  & 0.2 & 14.44 $\pm$ 0.24 & 2.08 $\pm$ 0.23 &  &  -  &  -  &  -  & 0 & 0.0281 $\pm$  0.0099 \\
3C409 & 63.4/-6.1 & 1.3 & 15.20 $\pm$ 0.04 & 1.13 $\pm$ 0.08 &  & 0.6 & 15.23 $\pm$ 0.01 & 1.02 $\pm$ 0.05 &  &  -  &  -  &  -  & 1 & 0.1479 $\pm$  0.0105 \\
3C410 & 69.2/-3.8 & 1.4 & 5.96 $\pm$ 0.01 & 0.81 $\pm$ 0.04 &  & 1.1 & 5.92 $\pm$ 0.01 & 0.52 $\pm$ 0.03 &  & 0.1 & 5.74 $\pm$    0.06 & 0.73 $\pm$ 0.26 & 0 & 0.7262 $\pm$  0.0545 \\
3C410 & 69.2/-3.8 & 0.2 & 10.97 $\pm$ 0.27 & 0.45 $\pm$ 0.64 &  &  -  &  -  &  -  &  &  -  &  -  &  -  & 1 & 0.0154 $\pm$  0.0219 \\
3C410 & 69.2/-3.8 & 0.2 & 11.54 $\pm$ 0.40 & 0.44 $\pm$ 0.50 &  &  -  &  -  &  -  &  &  -  &  -  &  -  & 2 & 0.0168 $\pm$  0.0192 \\
3C454.3 & 86.1/-38.2 & 0.9 & -9.41 $\pm$ 0.03 & 0.79 $\pm$ 0.07 &  & 0.2 & -9.62 $\pm$ 0.05 & 1.20 $\pm$ 0.14 &  &  -  &  -  &  -  & 0 & 0.0713 $\pm$  0.0063 \\
3C75 & 170.3/-44.9 & 1.6 & -10.28 $\pm$ 0.01 & 0.90 $\pm$ 0.03 &  & 1.1 & -10.29 $\pm$ 0.02 & 0.62 $\pm$ 0.05 &  &  -  &  -  &  -  & 0 & 0.1420 $\pm$  0.0042 \\
4C13.67 & 43.5/9.2 & 6.6 & 4.75 $\pm$ 0.01 & 1.54 $\pm$ 0.02 &  & 3.5 & 4.89 $\pm$ 0.01 & 1.65 $\pm$ 0.02 &  & 1.2 & 4.91 $\pm$    0.01 & 0.99 $\pm$ 0.03 & 0 & 15.6457 $\pm$  0.1521 \\
4C22.12 & 188.1/0.0 & 1.6 & -2.58 $\pm$ 0.16 & 0.94 $\pm$ 0.16 &  & 1.0 & -2.44 $\pm$ 0.01 & 1.19 $\pm$ 0.03 &  & 0.2 & -2.47 $\pm$    0.12 & 0.82 $\pm$ 0.21 & 0 & 1.9370 $\pm$  0.2582 \\
4C22.12 & 188.1/0.0 & 1.3 & -1.96 $\pm$ 0.16 & 0.67 $\pm$ 0.16 &  &  -  &  -  &  -  &  &  -  &  -  &  -  & 1 & 0.1639 $\pm$  0.0387 \\
G196.6+0.2 & 196.6/0.2 & 0.3 & 3.42 $\pm$ 0.14 & 2.14 $\pm$ 0.25 &  &  -  &  -  &  -  &  &  -  &  -  &  -  & 0 & 0.0575 $\pm$  0.0067 \\
G197.0+1.1 & 197.0/1.1 & 2.9 & 5.07 $\pm$ 0.16 & 1.93 $\pm$ 0.16 &  & 1.4 & 4.92 $\pm$ 0.03 & 1.90 $\pm$ 0.05 &  & 0.2 & 5.04 $\pm$    0.10 & 1.39 $\pm$ 0.24 & 0 & 2.2626 $\pm$  0.1892 \\
G197.0+1.1 & 197.0/1.1 & 0.8 & 7.18 $\pm$ 0.16 & 1.01 $\pm$ 0.16 &  & 0.6 & 7.33 $\pm$ 0.04 & 1.20 $\pm$ 0.08 &  &  -  &  -  &  -  & 1 & 0.0825 $\pm$  0.0130 \\
G197.0+1.1 & 197.0/1.1 & 0.4 & 16.34 $\pm$ 0.16 & 0.97 $\pm$ 0.16 &  & 0.0 & 16.63 $\pm$ 0.35 & 2.60 $\pm$ 0.35 &  &  -  &  -  &  -  & 2 & 0.0395 $\pm$  0.0065 \\
G197.0+1.1 & 197.0/1.1 & 0.4 & 17.39 $\pm$ 0.16 & 1.18 $\pm$ 0.16 &  & 0.1 & 17.52 $\pm$ 0.35 & 1.86 $\pm$ 0.35 &  &  -  &  -  &  -  & 3 & 0.0489 $\pm$  0.0066 \\
G197.0+1.1 & 197.0/1.1 & 0.5 & 32.24 $\pm$ 0.05 & 1.24 $\pm$ 0.02 &  & 0.3 & 32.28 $\pm$ 0.35 & 0.76 $\pm$ 0.35 &  &  -  &  -  &  -  & 4 & 0.0704 $\pm$  0.0010 \\
P0428+20 & 176.8/-18.6 &  -  &  -  &  -  &  &  -  &  -  &  -  &  &  -  &  -  &  -  & 0 &  -  \\
P0428+20 & 176.8/-18.6 & 0.9 & 10.71 $\pm$ 0.02 & 0.89 $\pm$ 0.05 &  & 0.3 & 10.59 $\pm$ 0.06 & 0.98 $\pm$ 0.14 &  &  -  &  -  &  -  & 1 & 0.0857 $\pm$  0.0049 \\
T0526+24 & 181.4/-5.2 & 8.0 & 6.86 $\pm$ 0.00 & 1.07 $\pm$ 0.01 &  & 6.3 & 6.90 $\pm$ 0.00 & 0.89 $\pm$ 0.01 &  & 0.5 & 6.95 $\pm$    0.02 & 0.83 $\pm$ 0.04 & 0 & 3.3663 $\pm$  0.0226 \\
T0629+10 & 201.5/0.5 &  -  &  -  &  -  &  &  -  &  -  &  -  &  &  -  &  -  &  -  & 0 &  -  \\
T0629+10 & 201.5/0.5 &  -  &  -  &  -  &  &  -  &  -  &  -  &  &  -  &  -  &  -  & 1 &  -  \\
T0629+10 & 201.5/0.5 & 1.2 & 1.13 $\pm$ 0.16 & 2.20 $\pm$ 0.16 &  & 0.5 & 1.30 $\pm$ 0.16 & 2.24 $\pm$ 0.16 &  &  -  &  -  &  -  & 2 & 0.2812 $\pm$  0.0203 \\
T0629+10 & 201.5/0.5 & 5.5 & 2.85 $\pm$ 0.16 & 1.25 $\pm$ 0.16 &  & 5.0 & 2.74 $\pm$ 0.16 & 1.15 $\pm$ 0.16 &  & 4.1 & 3.71 $\pm$    0.17 & 1.64 $\pm$ 0.17 & 3 & 59.8330 $\pm$  8.2582 \\
T0629+10 & 201.5/0.5 & 8.3 & 4.49 $\pm$ 0.16 & 1.85 $\pm$ 0.16 &  & 7.4 & 4.58 $\pm$ 0.16 & 1.63 $\pm$ 0.16 &  & 2.5 & 4.99 $\pm$    0.17 & 1.06 $\pm$ 0.17 & 4 & 36.1908 $\pm$  3.5195 \\
T0629+10 & 201.5/0.5 & 6.6 & 6.19 $\pm$ 0.16 & 1.40 $\pm$ 0.16 &  & 6.1 & 6.36 $\pm$ 0.16 & 1.33 $\pm$ 0.16 &  & 3.7 & 6.01 $\pm$    0.17 & 0.74 $\pm$ 0.17 & 5 & 51.6879 $\pm$  6.1885 \\
T0629+10 & 201.5/0.5 & 1.9 & 7.16 $\pm$ 0.16 & 1.57 $\pm$ 0.16 &  & 1.4 & 7.28 $\pm$ 0.16 & 1.39 $\pm$ 0.16 &  & 1.2 & 6.66 $\pm$    0.17 & 0.75 $\pm$ 0.17 & 6 & 21.8699 $\pm$  2.4927 \\
T0629+10 & 201.5/0.5 &  -  &  -  &  -  &  &  -  &  -  &  -  &  &  -  &  -  &  -  & 7 &  -  \\

\hline
\end{tabular}
\tablenotetext{1}{$T\rm_{peak}$:  Peak brightness temperature  corrected for main-beam efficiency of the telescope.}
\tablenotetext{2}{$O$: Order of the cold components along the line of sight, beginning with 0; larger numbers mean larger distances along the line of sight.}
\end{sidewaystable}



\end{document}